\newif\ifPDF
\documentclass[usenatbib]{mn2e}

\ifPDF
  \usepackage[T1]{fontenc}
  \usepackage{times}
  \usepackage[hypertex,colorlinks=true,citecolor=blue]{hyperref}
  \usepackage{aeguill}
  \usepackage[pdftex]{graphicx,color}
  \usepackage{subfigure}
  \usepackage{afterpage}
\else
  \usepackage[T1]{fontenc}
  \usepackage{times}
  \usepackage[dvips]{graphicx,color}
  \usepackage{subfigure}
  \usepackage{afterpage}
\fi

\title[Exploring the remarkable subpulse drift and polarization properties of PSR B0818$-$41]
{Exploring the remarkable subpulse drift and polarization properties of PSR B0818$-$41}
\author[B. Bhattacharyya, Y. Gupta and J. Gil]
{Bhaswati Bhattacharyya$^1$, Yashwant Gupta$^1$ and Janusz Gil$^2$\\\\
 $^1$National Centre for Radio Astrophysics, TIFR, Pune University Campus, Post Bag 3,
Pune 411 007, India\\
 $^2$Institute of Astronomy, University of Zielona Gora, Lubuska 2, 65-265 Zielona Gora, Poland }
\date{Accepted. Received}
\begin{document}
\label{firstpage}
\maketitle
\pagerange{\pageref{1}--\pageref{19}} \pubyear{2008}
\def\LaTeX{L\kern-.36em\raise.3ex\hbox{a}\kern-.15em
    T\kern-.1667em\lower.7ex\hbox{E}\kern-.125emX}

\begin{abstract}
PSR B0818$-$41 is one of the few pulsars which show multiple drift regions having well defined phase relationship.
In this paper we report new results from the multifrequency observations of this pulsar with the Giant Metrewave 
Radio Telescope (GMRT). We determine the mean flux of the pulsar at multiple frequencies and find evidence for a 
low frequency turn-over in its spectrum. Significant linear polarization is observed with depolarization at the 
edge of the profile, which is likely to be due to orthogonal polarization mode jumps. Circular polarization changes 
sign near the middle of the pulse profile at 1060 MHz which is not observed at 325 and 610 MHz. Remarkable frequency 
evolution of polarization angle (PA) is observed for this pulsar, which is commonly not seen in other pulsars. Based 
on the frequency evolution of average profile, observed PA swing and results from subpulse drifting, we propose two 
possible emission geometries, $\alpha\sim$11\degr, $\beta\sim-$5.4\degr and $\alpha\sim$175.4\degr, $\beta\sim-$6.9\degr. 
Simulation of the pulsar radiation pattern with both these geometries reproduces the observed features in the drift 
pattern quite well. In addition to the remarkable subpulse drifting observed at 325 MHz, we report subpulse drifting 
at 244 and 610 MHz. At 244 MHz we observe drifting in the outer regions but not in the inner region whereas at 610 MHz 
subpulse drift is observed in both inner and outer regions. Sometimes, we see changes of drift rates, transitions from 
negative to stationary and stationary to negative drift rates, many of which appear to have some connection with nulls. 
We determine $P_3^m$ (the measured time interval between the recurrence of successive drift bands at a given pulse 
longitude), $P_2^m$ (the measured longitude separation between two adjacent drift bands) and $\Delta \Phi$ (subpulse 
width) values at different frequencies for both inner and outer drift regions. Though $P_3^m$ is the same for all the 
drift regions, $P_2^m$ and $\Delta \Phi$ values are different. The peak emission from the leading and the trailing 
outer regions are offset by $\sim$ 9 $P_1$. Utilising this information, we solve the aliasing problem and argue that 
drifting is first order aliased with corresponding carousel rotation period $P_4\sim$ 10 s, which makes it the fastest 
known carousel. The drift pattern from the two rings are phase locked for PSR B0818$-$41. The same is found to be true for, 
all pulsars showing drifting in multiple rings of emission. This finding puts constraints on the theoretical 
models of pulsar emission mechanism and, favors a pan magnetospeheric radiation mechanism.
\end{abstract}

\section {Introduction}                                    \label{sec:intro}
Pulsars with drifting subpulses are considered as an important key for unlocking the mystery of how radio pulsars work. 
Though there has been significant progress both in high quality observations of drifting subpulses and in confronting 
their observational properties with existing models (e.g. \cite{Desh_etal}, \cite{Gupta_etal}, \cite{Weltevrede_etal_06}, 
\cite{Bhattacharyya_etal} (hereafter Paper 1)), the detailed nature of radio emission processes of pulsars and the exact 
location, distribution of the pulse emitting regions are still shrouded with mystery.

Knowledge of the viewing geometry is essential to understand the pulsar emission mechanism. The orientation of the pulsar 
radiation beam can be determined by two angles, $\alpha$ (the angle between the rotation axis and magnetic axis) and $\beta$ 
(the minimum angle between the magnetic axis and the observer's line of sight (LOS)). Possible geometry of a pulsar is 
determined by the "Rotating Vector Model (RVM)" (\cite{Radhakrishnan_etal}) fit to the polarization angle (PA) traverse. The 
possible choice of geometry should also satisfy the average profile evolution with frequency and the observed drift pattern. 

The phenomenon of subpulse drifting, first reported by \cite{Drake_etal}, is manifested as systematic subpulse behaviour: 
subpulses appear at progressively changing longitude in the pulse window following a particular path (known as drift band) 
which is specific to the individual pulsar concerned.
Studies of wide profile drifting pulsars can probe the emission regions in the pulsar magnetosphere (\cite{Gupta_etal}, Paper 1, 
\cite{Bhattacharyya_etal_08}). Since our LOS samples a large region of the polar cap in case of wide profile pulsars, presence 
of simultaneous multiple drift regions is quite probable. Phase relation between multiple drift regions are known only in few 
pulsars $-$ PSR B0815$+$09 \citep{McLaughlin_etal}, PSR B1839$-$04 \citep{Weltevrede_etal_06} and PSR B0818$-$41 (Paper 1). 
Observed phase relationship between the drift regions can probe the distribution and dynamics of the emission regions in the 
polar cap. Multifrequency observations may offer insights into various aspects of frequency dependence of pulsar radiation 
(\cite{Bhattacharyya_etal_08} and the references therein). 

PSR B0818$-$41 is a relatively less studied wide profile pulsar. In Paper 1, we reported the discovery of a remarkable drift 
pattern at 325 MHz, characterized by simultaneous occurrences of three drift regions: an inner region with flatter apparent 
drift rate flanked on each side by a region of steeper apparent drift rate. The drift regions have different values for $P_2^m$ 
(the measured longitude separation between two adjacent drift bands), but identical values for $P_3^m$ (the measured time 
interval between the recurrence of successive drift bands at a given pulse longitude). The closely spaced inner and outer 
drift regions always maintain a phase locked relationship (hereafter PLR). The observed drift pattern was interpreted as being 
created by the intersection of our LOS with two conal rings on the polar cap of a fairly aligned rotator ($\alpha\sim$11\degr), 
with an inner LOS geometry ($\beta\sim-$5.4\degr). Results from simulations reproduced the observed features (width of average 
profile, single pulse drift pattern etc). \cite{Qiao_etal} have studied the average polarization properties of this pulsar at 
660 and 1440 MHz. Their study reports a steep-sided wide profile with modest amount of linear polarization (32\% at 660 MHz 
and 31\% at 1440 MHz). From the wide integrated profile, \cite{Qiao_etal} also predicted that this pulsar must be a fairly aligned 
rotator.
 
We made multi epoch observations of PSR B0818$-$41 at 157, 244, 325, 610 and 1060 MHz using the GMRT in full polar mode. In Sect. 
\ref{sec:obs} we describe the observations. We determine pulsar flux at individual frequencies and investigate the spectrum in 
Sect. \ref{sec:fluxcal}. In Sect. \ref{sec:avp} we illustrate the average profile evolution with frequency and in Sect. \ref{sec:results_pol} 
we present a study of polarization properties of this pulsar. Possible choices of viewing geometries for PSR B0818$-$41 are discussed 
and compared with results from simulations of the radiation pattern in Sect. \ref{sec:viewing_geo}. Multi-frequency study of subpulse 
drifting is detailed in Sect. \ref{sec:results_drifting}. Unique phase relationships between multiple drift regions observed for PSR 
B0818$-$41 are investigated in Sect. \ref{sec:results_pecor}. In Sect. \ref{sec:Summary}, we present a final summary of different aspects 
of the analysis results and the interpretations.  
\begin{table}
\begin{center}
\begin{minipage}{80mm}
\caption{Details of the observations at different epochs and corresponding measurements of the values of mean fluxes}
\label{table:summary}
\begin{tabular}{l|c|c|c|c|c|c|c|c|c|c|c|c|c|c|c|c}
\hline
Frequency&Date of    &No of  &No of   &Mean Flux       \\
(MHz)    &observation&pulses &antennas& (mJy)          \\
         &           &       &used    &                                      \\
         &           &       &        &                                        \\\hline
157      &12/12/05   &2000   &16      & 87$^a$           \\
         &           &       &        & 105$^b$          \\\hline
157      &01/04/06   &3300   &17      & 69$^a$            \\\hline
157      &02/04/06   &4500   &19      & 67$^a$            \\\hline
244      &20/09/07   &9500   &16      & 46$^a$            \\
         &           &       &        & 62$^b$             \\\hline
244      &04/01/08   &9900   &16      & 53$^a$             \\
         &           &       &        & 74$^b$             \\\hline
244      &06/01/08   &4600   &16      & 64$^a$             \\
         &           &       &        & 108$^b$            \\\hline
325      &21/12/05   &6600   &15      & 95$^a$            \\
         &           &       &        & 116$^b$            \\\hline
325      &24/02/04   &3414   &17      & 93$^a$             \\\hline
325      &26/12/04   &1869   &19      & 65$^a$             \\\hline
325$^\dagger$ &10/03/06   &4300   &8  & 92$^a$             \\\hline
325$^\dagger$ &16/03/06   &6500   &7  & 98$^a$             \\\hline
610      &14/12/05   &9900   &16      & 20$^a$             \\
         &           &       &        & 20$^b$             \\\hline
610      &25/02/04   &1612   &20      & 29$^a$             \\\hline
610      &11/01/05   &3600   &22      & 24$^a$             \\\hline
610      &25/01/05   &7000   &20      & 19$^a$             \\\hline
610$^\dagger$ &10/03/06      &4300    &8& 17$^a$             \\\hline
610$^\dagger$ &16/03/06      &6500    &10& 21$^a$             \\\hline
1060     &22/11/05   &7800   &22         & 7$^a$              \\
         &           &       &           & 7$^b$              \\\hline
1060     &16/09/06   &6600   & 16        & 9$^a$              \\
         &           &       &           & 11$^b$             \\\hline
1060     &13/01/07   &3000   & 27        & 7$^a$              \\\hline

\end{tabular}

a : relative calibration\\
b : absolute calibration\\
$\dagger$ : simultaneous dual frequency observations \\
\end{minipage}
\end{center}
\end{table}
\section{Observations and preliminary analysis}                                   \label{sec:obs}
The GMRT is a multi element aperture synthesis telescope consisting of 30 antennas, each of 45 m diameter, spread over a region 
of 25 km diameter. The GMRT can also be used in array mode by adding the signals from the individual dishes, either coherently 
or incoherently. We performed polarimetric observations of PSR B0818$-$41 at multiple frequencies (157, 244, 325, 610 and 1060 MHz), 
with the GMRT, using the phased array mode (\cite{Gupta_etal_00}), at several epochs. For the polarimetric observations we used the 
technique described by \cite{Mitra_etal_07}. The simultaneous dual frequency observations reported in this paper were done using the 
"band masking" technique described in \cite{Bhattacharyya_etal_08}. The signals from different antennas in the observing frequency 
bands were eventually converted to a base-band signal of 16 MHz band-width, which was sampled at the Nyquist rate. The 16 MHz observing 
band-width was divided into 256 spectral channels in the FX correlator. In the GMRT array combiner (GAC), the signals from selected 
antennas were added coherently to get the phased array output for each polarization which was then fed to the pulsar receiver. The 
pulsar receiver computes both the auto and cross polarized powers for each of the 256 channels. These data were further integrated in 
time and finally recorded to disk at a sampling rate of 0.512 ms. A suitable calibration procedure as described by \cite{Mitra_etal} 
was applied to the recorded data to recover the stokes parameters  I, Q, U, V. During the off-line analysis the raw data were further 
integrated to achieve the final resolution of 2.048 ms. Table \ref{table:summary} summarises the observations.

Average pulse profiles at different observing frequencies were obtained by dedispersion with a DM of 113.4 $pc/cm^3$. Power line (50 Hz) 
and its harmonics were the major contributors to the radio frequency interferences present in the data. To remove power line related 
interferences, the dedispersed data were put through a radio frequency interference filtering routine which detected most (but probably 
not all) of the power line interferences and replaced them by appropriate random noise. The dedispersed, interference free data stretch 
were synchronously folded with the topocentric pulsar period.

\begin{figure}
\includegraphics[angle=-90, width=0.5\textwidth]{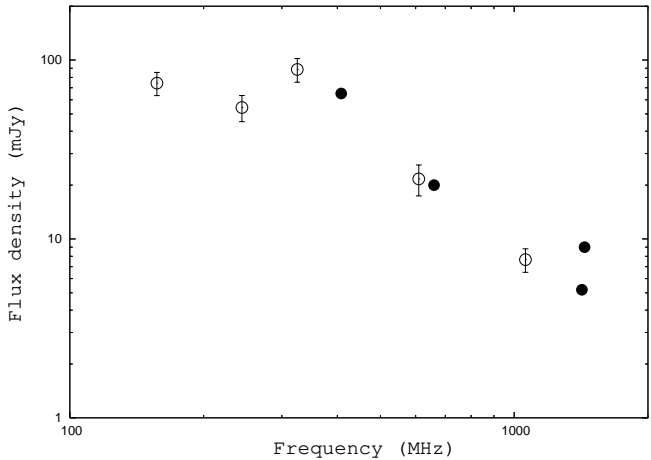}
\caption{Spectra with the calculated mean flux densities from relative calibration (Table \ref{table:summary}). Open circles with error bars represent the flux measurements from this work and the filled circles represents the flux values available in the literature. Spectra flattens below 325 MHz, at frequencies like 157 and 244 MHz. Known values of the mean flux of PSR B0818$-$41 at different frequencies are: 65 mJy at 400 MHz (Hobbs et al 2004), 20 mJy at 660 MHz (Qiao et al. 1995), 5.2 mJy and 9 mJy at 1.4 GHz estimated by Taylor et al. (1993) and Qiao et al. (1995) respectively. The spectra from our work seems to disagree with mean flux of 9 mJy at 1.4 GHz (Qiao et al. 1995) and agrees with mean flux of 5.2 mJy at 1.4 GHz (Taylor et al. 1993).}
\label{fig:sp}
\end{figure}
\begin{figure*}
\begin{minipage}{180mm}
\begin{center}
\includegraphics[angle=0,width=0.6\textwidth]{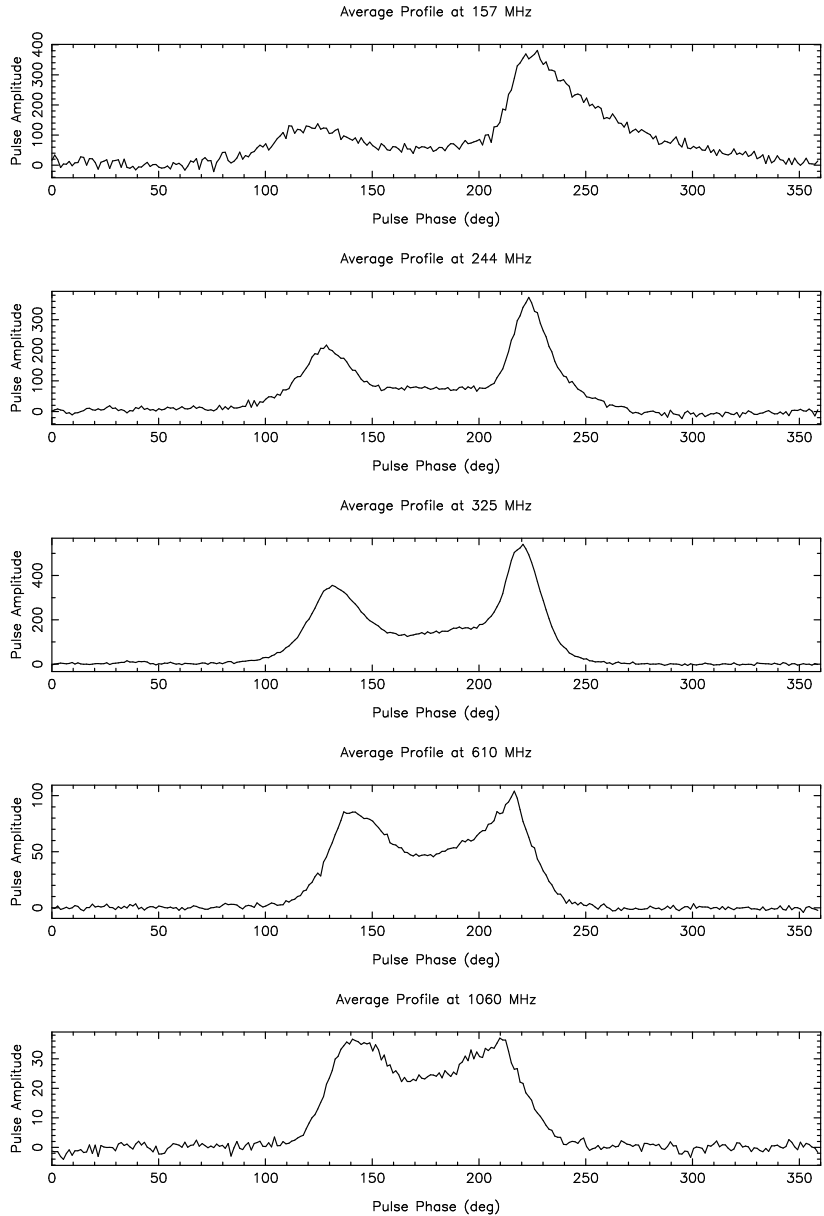}
\caption{Frequency evolution of the average profile of PSR B0818$-$41. Different panels plot the average profiles obtained from the GMRT observations at different observing frequencies, 157, 244, 325, 610 and 1060 MHz respectively. Pulse amplitude is in mJy. The separation between two peaks of the double peaked pulse profile decreases with increasing frequency following the radius to frequency mapping.}
\label{avp_157_244_325_610_1060}
\end{center}
\end{minipage}
\end{figure*}
\section{Flux at individual frequencies and spectrum}                      \label{sec:fluxcal}
We determine the mean flux of PSR B0818$-$41 for all the observing frequencies at all the observing epochs. For some of the epochs 
we had observed flux calibrator sources and for these cases we determine the flux from absolute calibration. The calibrator fluxes 
at 90 and 20 cm available in the VLA catalog are used as the calibrator fluxes at 325 and 1060 MHz respectively. However, the calibrator 
flux at 610 MHz is obtained by linear interpolation between 90 and 20 cm fluxes from the VLA catalog and the calibrator fluxes at 157 
and 244 MHz are obtained by linear extrapolation of the 90 and 20 cm fluxes. For the epochs where we do not have any flux calibrator 
source observed, we follow a relative calibration procedure, using the knowledge of the observing parameters and the background sky 
temperature near the source (similar to as done in \cite{Bhattacharyya_etal_08}). As a crosscheck we compare the fluxes where both 
absolute and relative calibrations were possible. The flux values determined from absolute and relative calibration are similar for 
all the frequencies, except at 157 and 244 MHz. The reason for the difference in flux values between absolute and relative calibration 
could be due to the error in the estimation of calibrator flux or inaccurate knowledge of the sky temperature at the position of the 
pulsar, at the lower frequencies. In Table \ref{table:summary} we list the mean flux, for all the five frequencies of observations. 
\begin{table*}
\begin{minipage}{180mm}
\caption{Frequency dependence of $\Delta\Phi$, $P_3^m$, $P_2^m$ and $\Delta\Phi_s$}
\label{table2}
\begin{tabular}{|c|c|c|c|c|c|c|c|c|c|c|c|c|c|}
\hline
Frequency   & \multicolumn {3}{|c|}{$\Delta\Phi$} & $P_3^m$      &\multicolumn{3}{|c|}{$({P_2^m})^c$}&\multicolumn {3}{|c|}{$(\Delta\Phi_s)^{c}$}\\
            & \multicolumn {3}{|c|} {(\degr)}     &                     & \multicolumn {3}{|c|} {(\degr)} & \multicolumn {3}{|c|} {(\degr)}   \\
(MHz)       & observation  &  simulation$^a$ & simulation$^b$&             &inner&trailing outer&leading outer&inner&trailing outer&leading outer \\
            &              &                 &               &                   &       &             &            &       &             &    \\\hline
157         & 105$\pm$0.3  &      104        & 104           & 18.6$\pm$1.6$P_1$ &  $-$  &   $-$       &  $-$       & $-$   &   $-$       &  $-$\\
            &              &                 &               &                   &       &             &            &       &             &    \\
244         & 96$\pm$0.3   &      93         & 94            & 18.6$\pm$1.6$P_1$ &  $-$  &  16.9       &  29.8      & $-$   & 13.5        &13.3 \\
            &              &                 &               &                   &       &             &            &       &             &\\
325         & 86.6$\pm$0.3 &      88         & 88            & 18.3$\pm$1.4$P_1$ & 17.6  &  21.5       &  31        & 6.3   & 10.8        &12.3 \\
            &              &                 &               &                   &       &             &            &       &             &\\
610         & 75.7$\pm$0.3 &      74.6       & 75            & 18.6$\pm$1.5$P_1$ & 16.5  &   $-$       &  $-$       & 4.3   & 10.8        &14.1 \\
            &              &                 &               &                   &       &             &            &       &             &\\
1060        & 63$\pm$0.3   &      64         & 65            &       $-$         & $-$   &   $-$       &  $-$       & $-$   &  $-$        & $-$\\\hline
\end{tabular}
\\
$a$ : simulation is done with {\bf G-1} ($\alpha=$11\degr and $\beta=-$5.4\degr), with emission height formula $r_{em}=41 R \nu^{-0.24} {P_{-15}}^{0.063}{P_1}^{0.26}$\\
$b$ : simulation is done with {\bf G-2} ($\alpha=$175.4\degr and $\beta=-$6.9\degr), with emission height formula $r_{em}=41 R \nu^{-0.18} {P_{-15}}^{0.05}{P_1}^{0.3}$\\
$c$ : error associated in determination of $\Delta\Phi_s$ is $\pm$1.3 \degr
\end{minipage}
\end{table*}
Though the scintillation properties of PSR B0818$-$41 are not studied, still we try to understand the effect of scintillation on the 
observed flux, by comparing with two nearby pulsars for which the scintillation properties are known. The mean flux densities for two 
nearby pulsars, B0628$-$28 and PSR B0834+06, are observed to change by $\sim$ 30\% between the observing epochs \citep{Bhat_etal_99}. 
Between different observing sessions of PSR B0818$-$41, we observe 15 to 20\% change of the mean flux, for a particular frequency 
(Table \ref{table:summary}). Hence, it is likely that the estimated mean flux values of PSR B0818$-$41 are affected by the scintillation 
to a moderate extent. Fig. \ref{fig:sp} presents the spectral behaviour of the mean flux densities obtained from relative calibration. 
The error in mean flux is calculated from the standard deviation of the available flux values at a given frequency. Our estimation of 
the mean flux values of PSR B0818$-$41 at different frequencies are in the same ballpark as the earlier results available at 400, 660 
and 1400 MHz (Fig. \ref{fig:sp}). Monotonic increase in pulsar flux is observed with decreasing frequency starting from 1400 to 325 MHz 
with average value of spectral index $\alpha \sim-2$ (where $\alpha$ is defined as $I \propto \nu^\alpha$), which is similar to the mean 
value of the power law index for a typical pulsar spectrum \citep{Maron_etal}. However, we note that there is a tendency of flattening of 
the pulsar spectrum at frequencies lower than 325 MHz (at 244 or 157 MHz). We believe that this flattening of the spectrum is associated 
with the turn-over at low frequencies. But because of the uncertainty in the observed flux values, it is difficult to estimate the exact 
frequency of the turn-over. Usually pulsars exhibit such a power law spectrum with a low frequency turn-over near $\sim$ 100 MHz 
\citep{Malofeev_etal}. For PSR B0818$-$41, the flux of leading outer, trailing outer and inner region evolve differently with frequency 
(Fig. \ref{avp_157_244_325_610_1060}). Relative strength of the leading peak as compared to the trailing peak increases with frequency. 
Inner saddle gets more filled with increasing frequency which is evident from the average profiles plotted in Fig. \ref{avp_157_244_325_610_1060}. 
Hence, spectral indices of different profile components are different for PSR B0818$-$41. Component dependent frequency evolution of flux 
is observed for many other pulsars (e.g. for PSR B1133$+$16 reported by \citep{Bhat_etal}).
\section{Average profile evolution with frequency}                                  \label{sec:avp}
Fig. \ref{avp_157_244_325_610_1060} presents the plot of the average profiles at five different observing frequencies. We observe a steep 
sided double peaked average profile at 325 MHz, as reported in Paper 1. At 157 MHz, the trailing peak is broadened, probably because of 
scattering effects, which become more prominent at lower frequencies. The two peaks of the average profile, at 157 MHz, are of unequal 
intensity, leading peak having about half the intensity of the trailing peak. With increase in frequency, the leading peak becomes more 
intense and the saddle region joining the two peaks becomes more filled up. At 1060 MHz, both the peaks of the double peaked profile are 
of comparable intensity. Separation between the two peaks of the double peaked average profile ($\Delta \Phi$) reduces with increasing 
frequency (Table \ref{table2}). $\Delta \Phi$ varies with frequency as $\Delta \Phi \propto \nu^{-p}$, with average $p$ value of 0.2 
(Table \ref{table2}). This follows the general trend of the "radius to frequency mapping", seen in a majority of pulsars. For another wide 
profile pulsar, PSR B0826$-$34, we reported similar frequency evolution of $\Delta \Phi$ \citep{Bhattacharyya_etal_08}. Although the 
simulations described in Sect. \ref{sec:sim} with both the geometries ({\bf G-1} and {\bf G-2}) broadly reproduce the evolution of 
$\Delta \Phi$ with frequency, it is hard to reproduce the observed intensity difference between the leading and trailing peaks.
\section{Polarization study}                            \label{sec:results_pol}
Fig. \ref{fig:pol_325}, \ref{fig:pol_610} and \ref{fig:pol_1060} present the polarization profiles at 325, 610 and 1060 MHz. 
Polarization profiles at 610 and 1060 MHz have better signal to noise than the published polarization profiles at 660 and 
1440 MHz by \cite{Qiao_etal}, and reproduce all the basic polarization features reported by them. We have observed PSR 
B0818$-$41 at several epochs. Analysis of the data from all the epochs produces self consistent polarization profiles. The 
polarization profiles with best signal to noise ratio available for a given frequency are presented in this work. 

\subsection {Linear polarization}
Significant amount of linear polarization is observed at all the frequencies (about 45\% at 325 MHz, 32\% at 610 MHz and
20\% at 1060 MHz) : the percentage of linear polarization is more at 325 than at 610 and 1060 MHz. This agrees well
with the percentage of linear polarization quoted by \cite{Qiao_etal}. The average profile for linear polarization
follows the general double peaked structure of the total intensity profile. But the separations between the two peaks
of the linearly polarized profile are, 79\degr at 325 MHz, 71\degr at 610 MHz and 60\degr at 1060 MHz, which are less
than the separations between the total intensity peaks at individual frequencies. We note that this phenomenon appears
to have a frequency dependence: width of linearly polarized and the total intensity profile differing
by $\sim$10\% at 325 MHz whereas by $\sim$5\% at 610 and 1060 MHz. At 325 MHz the leading and the trailing peaks of the
average linear polarization profile are of similar intensity, whereas in the total intensity profile the trailing peak
is significantly more intense than that of the leading one. At 610 and 1060 MHz the leading and the trailing peaks of
the linear polarization profile are of similar intensity, and the same is true for the total intensity profiles. Average
linear polarization falls off much faster than the total intensity and decreases to zero near the outer edges of the
profiles. This effect of edge depolarization is less in 1060 MHz as compared to 325 and 610 MHz.
\begin{figure}
\begin{center}
\includegraphics[angle=-90, width=0.5\textwidth]{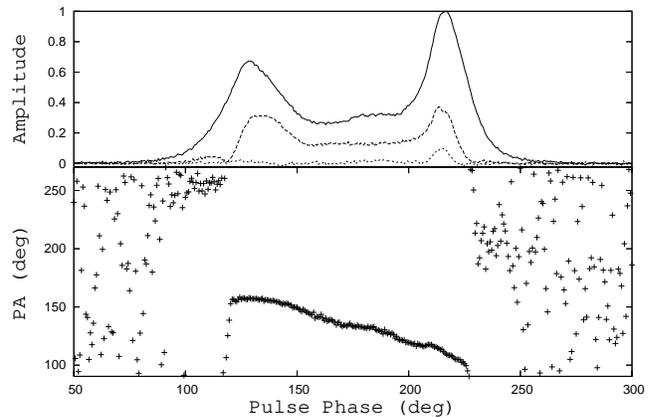}
\caption{Polarization profile of PSR B0818$-$41 at 325 MHz. The top panel shows the average profile in total intensity (solid line), linear polarization (dashed line) and circular polarization (dotted line). The lower panel shows the variation of PA as a function of longitude. The amplitude is in arbitrary units.}
\label{fig:pol_325}
\end{center}
\end{figure}
\begin{figure}
\begin{center}
\includegraphics[angle=-90, width=0.5\textwidth]{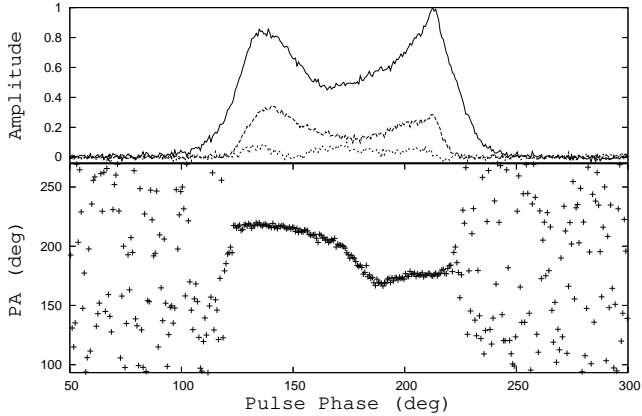}
\caption{Same as Fig. \ref{fig:pol_325}, but at 610 MHz}
\label{fig:pol_610}
\end{center}
\end{figure}
\begin{figure}
\begin{center}
\includegraphics[angle=-90, width=0.5\textwidth]{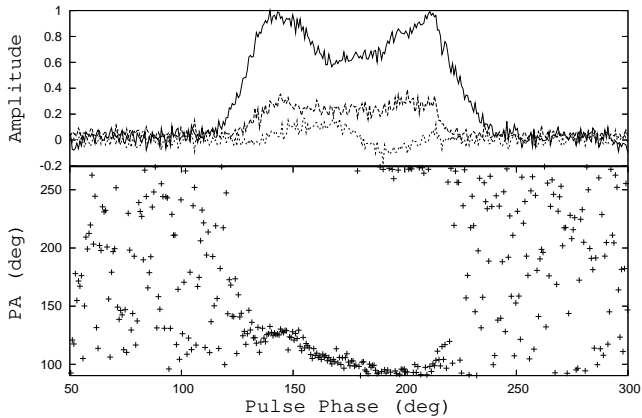}
\caption{Same as Fig. \ref{fig:pol_325}, but at 1060 MHz}
\label{fig:pol_1060}
\end{center}
\end{figure}
One of the striking features seen in Fig. \ref{fig:pol_325}, \ref{fig:pol_610} and \ref{fig:pol_1060} is that the PA curve 
evolves significantly with frequency. In order to do a closer comparison, Fig. \ref{pa_325_610_1060} shows the three PA 
curves in one plot. The PA curve shows linear monotonic variations (with some amount of irregularities) at 325 MHz. Total 
PA swing is about 69\degr at 325 and about 53\degr at 610 MHz. The PA curve at 610 MHz is quite different than that of 325 MHz. 
For the left half of the PA curve at 610 MHz, nearly linear PA variation with pulse longitude is observed, which is similar to 
325 MHz. A different slope of the PA curve is seen at 610 MHz at the middle of the pulse profile. The other part of the PA curve 
at 610 MHz exhibits kinky structure and some signature of reversal of the slope of the PA curve is observed at the edge of the 
profile. It is difficult to explain the total PA sweep at 610 MHz with the RVM \citep{Radhakrishnan_etal}, which is indicative 
of non dipolar emission geometry. Though the PA curve is quite noisy at 1060 MHz, we see a behaviour intermediate to that at 325 
and 610 MHz (Fig. \ref{pa_325_610_1060}).

Evidence of orthogonal polarization mode jump is observed at the two edges of the profile at 325 MHz (Fig. \ref{fig:pol_325}).
While investigating the single pulse PA using scatter plot, we observe orthogonal polarization mode jump near two edges of the 
pulse profile at both 325 and 610 MHz. In 1060 MHz the signal to noise was not enough to observe single pulse PA. Orthogonal 
polarization mode jump observed at the edge of the profile (at 325 and 610 MHz), can be the reason of the abrupt decrease of 
the linear polarization near the edge of profile. Presence of the secondary polarization mode at the edge of the profile can 
cause the observed reduced width of the linearly polarized profile compared to the total intensity profile. Stronger orthogonal 
polarization mode jump observed at 325 MHz than 610 or 1060 MHz, can explain the fact that observed differences between the 
separations between the peaks of linearly polarized and total intensity profile are more at 325 MHz. For another wide profile 
pulsar PSR B0826$-$34 we observed similar orthogonal polarization mode change at the edge of the profile, resulting in sharp 
fall of the linear polarization \citep{Biggs_etal}.
\subsection{Circular polarization}
Though circular polarization observed for PSR B0818$-$41 at 325 MHz is quite less, under the trailing component of the pulse 
profile there is a feature of increased circular polarization. Small amount of circular polarization is observed at 610 and 
1060 MHz. Circular polarization changes sign at 1060 MHz near the middle of the pulse profile, which is also reported by 
\cite{Qiao_etal} at 1440 MHz. We note that changing sign of circular polarization is not observed at 325, 610 (this work) 
and 660 MHz \citep{Qiao_etal}. 

There are two possible origins of circular polarization of pulsars: either intrinsic to the emission properties and dependent 
on the emission mechanism (e.g. \cite{Michel}, \cite{Radhakrishnan_90_etal}, \cite{Gangadhara}) or generated by propagation effects 
(e.g. \cite{Cheng_etal}). Circular polarization of some pulsars changes with frequency and the variation of degree of circular 
polarization with frequency is very different from pulsar to pulsar (e.g. refer to Table 5 of \cite{Han_etal}). PSR J2053$-$7200 
is an extreme example where sense reversal of circular polarization from the positive to negative direction is observed near the 
intersection of two components, at low frequencies, whereas sense of circular polarization changes from negative to positive 
direction at high frequencies \citep{Qiao_etal}. 

The feature of increased circular polarization near the trailing peak at 325 MHz, which is not observed at 610 and 1060 MHz, may 
contribute towards the more intense trailing peak than the leading one observed at 325 MHz.

Sense reversals of the circular polarization are commonly associated with the core components. However, for PSR B0818$-$41, at 
1060 MHz, it is associated with the conal components (Paper 1). There are some conal double pulsars such as PSR B0826$-$34, PSR 
B2048$-$72 which show a central reversal of circular PA but no core component at all (\cite{Biggs_etal}, \cite{Han_etal}). In 
such cases the sense reversal is associated with conal components rather than the core components \citep{Biggs_etal}.

For PSR B0818$-$41 we observe a correlation between the sign change of circular polarization and the sense of PA sweep: with right 
hand (negative) circular polarization accompanying increasing PA. \cite{Han_etal} reported similar correlation between the sense of 
PA sweep and the sense of circular polarization, using the polarization data of 20 conal-double pulsars.
\subsection {RVM fit to PA sweep :}    \label{sec:viewing_geo}
RVM \citep{Radhakrishnan_etal} is fitted to the PA curve of PSR B0818$-$41 at 325 MHz. While fitting we used symmetry about 
magnetic axis, i.e. $0<\alpha<180$\degr (e.g \cite{Everett_etal} and \cite{Johnston_etal}), rather than symmetry about the magnetic 
equator, i.e. $0<\alpha<90$\degr (e.g. \cite{Lyne_etal}, \cite{Narayan_etal}, \cite{Rankin_90}, \cite{Rankin_93} etc). Our attempt 
to fit RVM to the PA curve at 325 MHz gives best fit values of $\alpha \sim 175.4$\degr and $\beta\sim-6.9$\degr, which is an outer 
LOS geometry. This means that the spin axis is anti aligned to the magnetic axis. The angle between the spin axis and the magnetic axis is, 
180$-$175.4$\sim$5.6\degr, and the LOS makes a circle of radius $5.6+\beta$=12\degr around the rotation axis. Fig. \ref{pa} presents 
a plot of the observed PA sweep at 325 and 610 MHz and the RVM curves for two possible geometries : $\alpha \sim 11$\degr, $\beta \sim -5.4$\degr 
(the geometry described in Paper 1 which incidentally reproduces the middle part of PA sweep at 610 MHz) and $\alpha \sim 175.4$\degr, 
$\beta \sim -6.9$\degr (the geometry derived from the RVM fit to 325 MHz PA sweep). 
\section{Understanding the emission geometry of PSR B0818$-$41}           \label{sec:viewing_geo}
\begin{center}
\begin{figure}
\begin{center}
\includegraphics[angle=-90,width=0.5\textwidth]{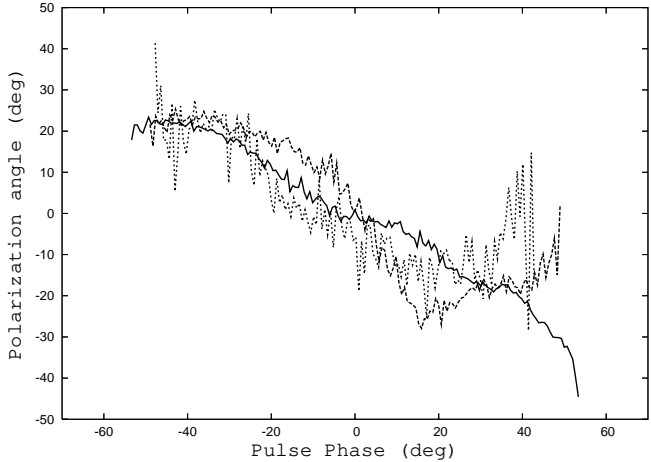}
\caption{Observed PA curves at 325 (solid), 610 (doted line), 1060 (dash-dot)}
\label{pa_325_610_1060}
\end{center}
\end{figure}
\end{center}
\begin{center}
\begin{figure}
\begin{center}
\includegraphics[angle=-90,width=0.5\textwidth]{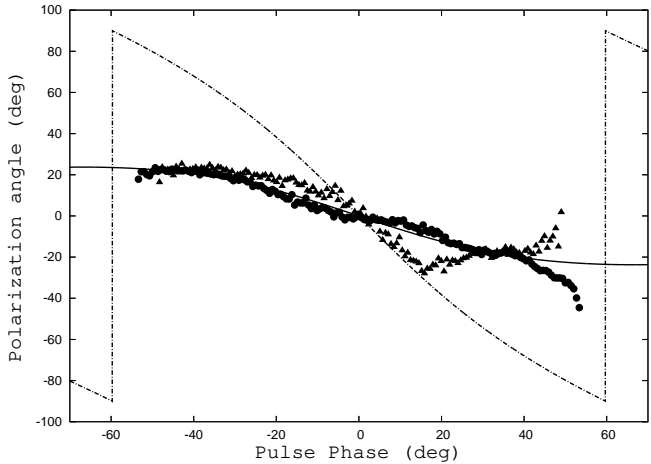}
\caption{Observed PA curves at 325 (solid circles), 610 MHz (solid triangles) and the RVM curve for $\alpha\sim175.4$\degr and $\beta\sim-6.9$\degr (dotted line) and for $\alpha\sim11$\degr, $\beta\sim-5.4$\degr (dashed line)}
\label{pa}
\end{center}
\end{figure}
\end{center}
The possible choices of the emission geometries are constrained by the following observed characteristics:\\
(i) Observed profile width at individual frequencies (Table \ref{table2}) as well as the profile evolution with frequency 
(shown in Fig. \ref{avp_157_244_325_610_1060}). \\
(ii) Overall pattern of drifting with simultaneous multiple drift regions (e.g. Fig. \ref{sp_325}), $P_2^m$, $P_3^m$  values 
and the relative intensities of the drift regions. \\
(iii) Observed PA curves.

As described in Paper 1 (see Sect. \ref{sec:intro}), the observed drift pattern for PSR B0818$-$41 can be interpreted 
as being created by the intersection of our LOS with two conal rings on the polar cap of a fairly aligned rotator 
($\alpha\sim$11\degr), with an inner LOS geometry ($\beta\sim-$5.4\degr). This geometry was developed to give best fit 
to profile shape and drift pattern at 325 MHz. In this work we find that the PA traverse at 325 MHz can not be explained by 
this geometry, but surprisingly the middle part of the PA swing at 610 MHz is satisfied by this geometry (Fig. \ref{pa}). 
However, best RVM fit to the PA curve at 325 MHz as well as left half of PA sweep at 610 MHz is obtained with $\alpha\sim$175.4\degr 
and $\beta\sim-$6.9\degr (Sect. \ref{sec:results_pol}). Therefore, we have two competing geometries, \\
{\bf G-1} ($\alpha=$11\degr and $\beta=-$5.4\degr) : geometry used in Paper 1, that gives best fit to profile shape and drift 
pattern at 325 MHz. This geometry incidentally produces reasonable fit to the middle part of PA sweep at 610 MHz. \\
{\bf G-2} ($\alpha=$175.4\degr and $\beta=-$6.9\degr) : geometry obtained by RVM fitting to the PA curve at 325 MHz.\\
In order to investigate how well these two geometries satisfy the above constraints we have exhaustively used the simulation 
techniques as described below.\\

\subsection{Simulation}                                        \label{sec:sim}
The simulations with two rings of emission having same carousel rotation period ($P_4$) and the number of sparks ($N_{sp}$), with 
180\degr out of phase emission between the rings, successfully reproduces the PLR in the observed drift pattern. Considering the 
measured $P_2^m=$17.5\degr for the inner region and assuming $P_2^m\sim P_2^t$ (true longitude separation between adjacent drift bands) 
for nearly aligned rotator, we have, $N_{sp}\sim19-22$. We simulate the radiation from PSR B0818$-$41 using the method described 
in \cite{Gil_etal_00} for both {\bf G-1} and {\bf G-2}. While doing the simulation we use 20 equispaced sparks in each of the two 
concentric rings, rotating with a drift rate of 19.05\degr/$P_1$ (as $P_3^m\sim 18P_1$, Sect. \ref{sec:results_drifting}). The empirical 
emission altitude formula by \cite{Kijak_etal} is used for determining the LOS cut through the beam at different frequencies. 

\begin{figure}
\begin{center}
\includegraphics[angle=0, width=0.44\textwidth]{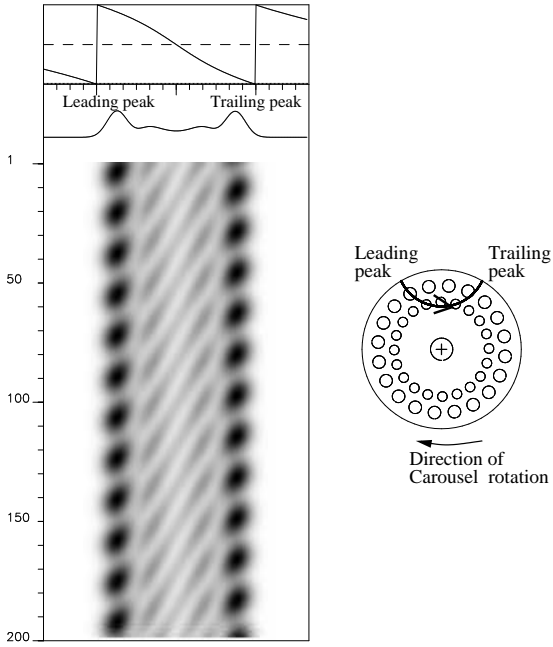}
\caption{Simulation of the subpulse drift pattern with simple dipolar geometry for the case: $\alpha=11$\degr, $\beta=-5.4$\degr, drift rate $D=19.05\degr/P_1$, $N{sp}=20$. Corresponding $P_5=8.7P_1$.}
\label{fig:sim1}
\end{center}
\end{figure}

\begin{figure}
\begin{center}
\includegraphics[angle=-90, width=0.44\textwidth]{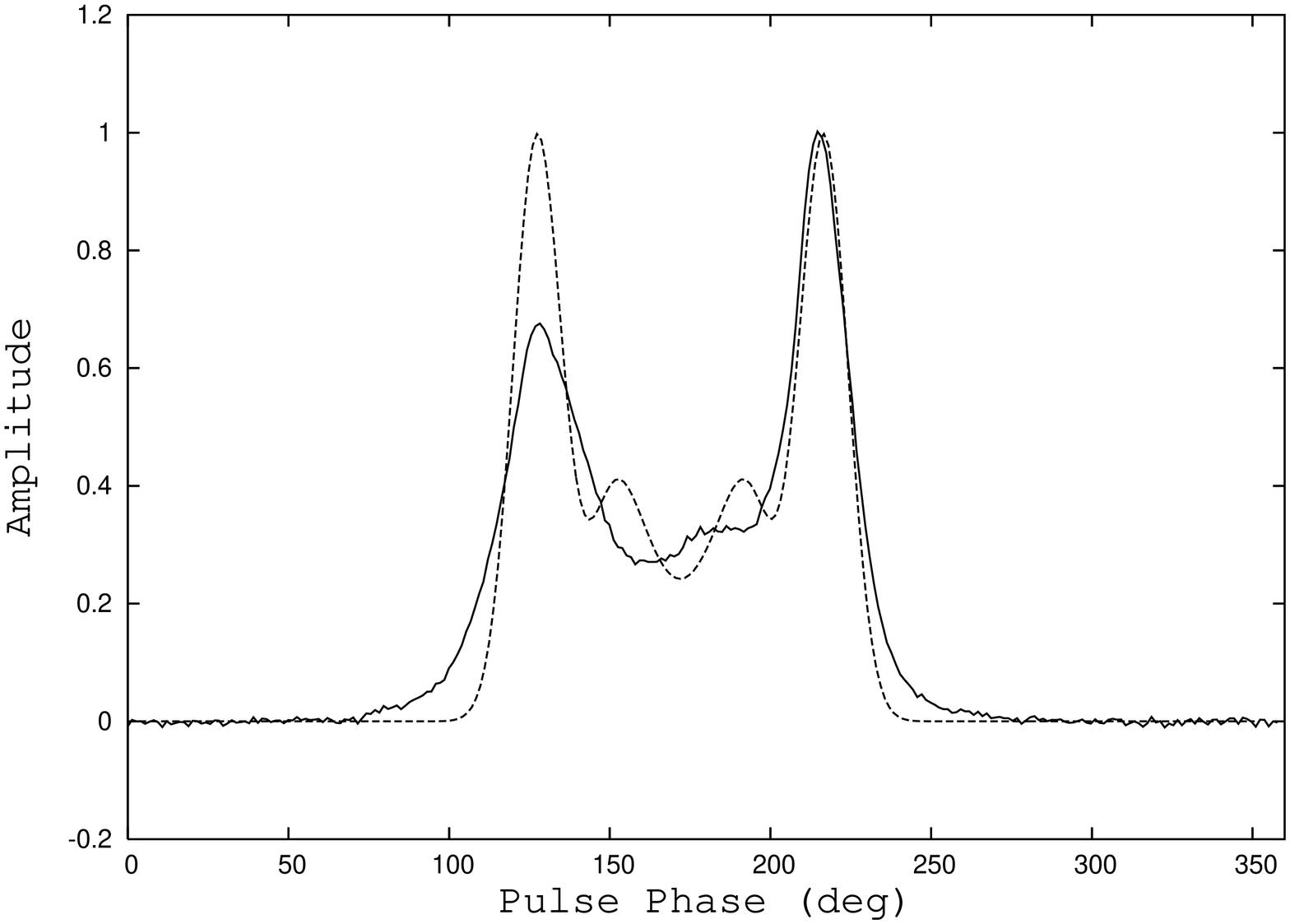}
\caption{Observed and simulated pulse profile at 325 MHz with simple dipolar geometry for the case: $\alpha=11$\degr, $\beta=-5.4$\degr; drift rate$D=19.05\degr/P_1$.}
\label{fig:prof_sim1}
\end{center}
\end{figure}

\begin{figure}
\begin{center}
\includegraphics[angle=0, width=0.44\textwidth]{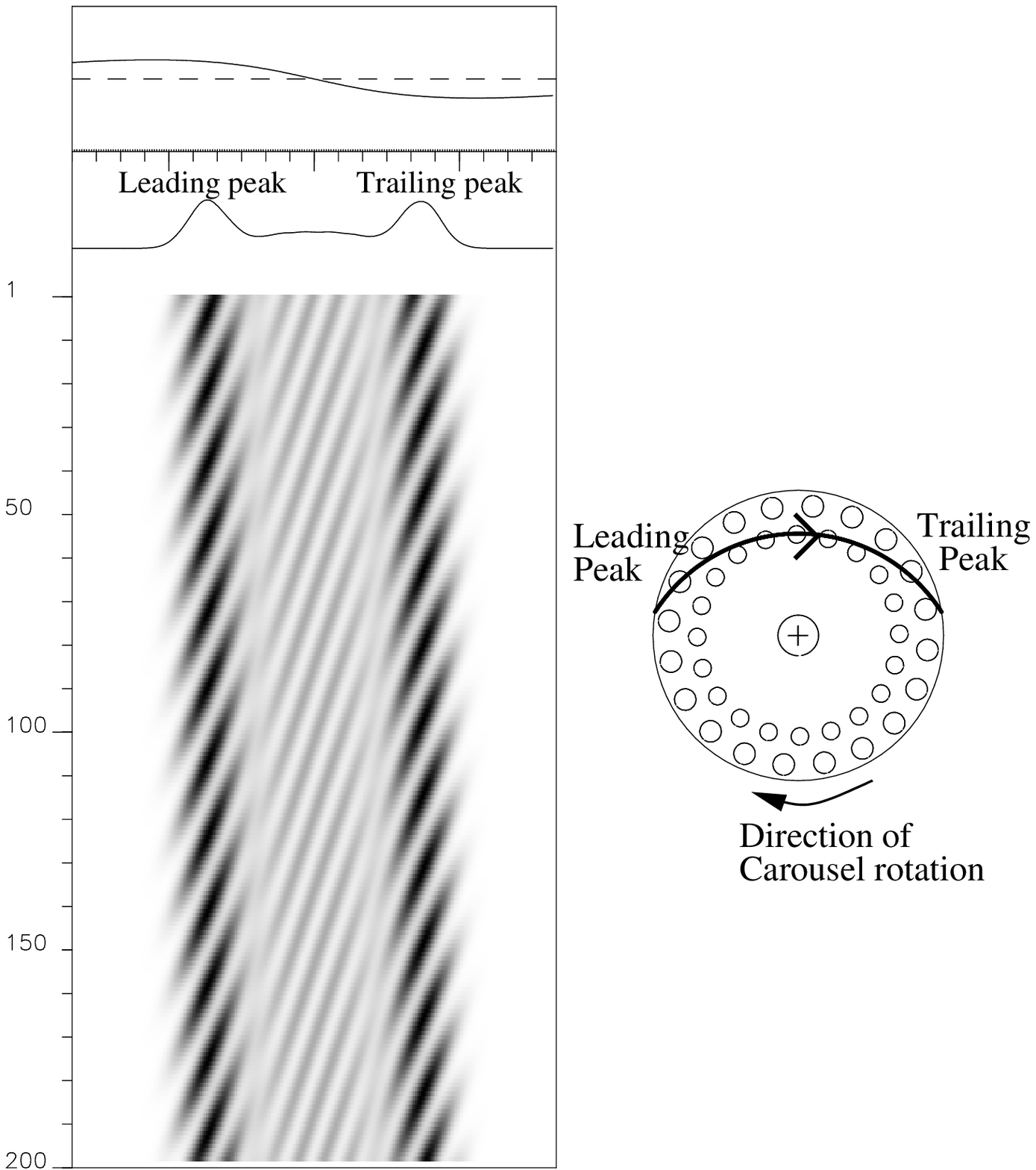}
\caption{Simulation of the subpulse drift pattern with simple dipolar geometry for the case: $\alpha=175.4$\degr, $\beta=-6.9$\degr, drift rate $D=19.05\degr/P_1$, $N{sp}=20$, Corresponding $P_5=8.7P_1$.}
\label{fig:sim}
\end{center}
\end{figure}

The gray scale plots of the simulated single pulses for {\bf G-1} and {\bf G-2} (Fig.\ref{fig:sim1} and Fig.\ref{fig:sim}) reproduces 
the overall pattern of drifting. Radii of the inner and outer rings in Fig.\ref{fig:sim1} (simulated with {\bf G-1}) are 0.6 $r_p$ and 
0.8 $r_p$, which are identical to the values used in Paper 1. However, for Fig. \ref{fig:sim} (simulated with {\bf G-2}) radii of the 
inner and outer rings are 0.7 $r_p$ and 0.9 $r_p$ respectively. Ratio of spark intensities of the inner and outer rings, required to 
reproduce the relative intensities of the inner saddle region and the trailing peak, is $\sim$ 3. This is identical to the relative 
intensities of the inner saddle region and the trailing peak of the observed pulse profile at 325 MHz (Fig. \ref{avp_157_244_325_610_1060}). 
From the simulated single pulses with {\bf G-1}, we calculate $P_2^m\sim$ 16\degr for both inner and outer regions, which is close to 
the observed $P_2^m$ in the inner region ($P_2^m\sim17.5\degr$ in Sec. \ref{sec:P2_det}) whereas for {\bf G-2}, $P_2^m\sim$ 14\degr for 
the inner and outer regions, which is a bit lower than the observed $P_2^m$ for the inner region. Fluctuation spectrum of the single 
pulses generated from both the geometries exhibits a strong feature at $\sim18 P_1$, which is identical to the determined $P_3^m$ value. 
Simulated single pulse drift pattern with {\bf G-1} in Fig.\ref{fig:sim1} have single drift bands in the outer drift region on both sides 
of the inner region, which matches with observation. But for the simulated single pulse drift pattern with {\bf G-2}, at times, two drift 
bands in the outer region on both sides of the inner region are observed. The simulated average profiles for both the geometries and 
the observed average profiles at 325 MHz are over plotted in Fig. \ref{fig:prof_sim1}. In the simulated average profiles the inner saddle 
region get more filled at higher frequencies, which reproduces the observations (Fig. \ref{avp_157_244_325_610_1060}). Though, general 
double peaked structure of the profiles, profile width at individual frequencies as well as the profile evolution with frequency are well 
reproduced by the simulations with both the geometries (Table \ref{table2}), features like differences in amplitude of the leading and 
trailing peaks are not achieved with simulations. 

To summarise, we have simulated the radiation pattern of the pulsar for both {\bf G-1} and {\bf G-2}. Both the geometry satisfy constraints 
(i), (ii) and (iii) discussed in Sect. \ref{sec:viewing_geo}. Though the overall PA curve at 325 MHz, and left half of the PA curve at 610 
MHz is fitted with {\bf G-2}, only the middle part of the PA curve at 610 MHz is fitted with {\bf G-1}. On the other hand, {\bf G-1} appears 
to give a better match to the observed $P_2^m$ values in the inner region. We will refer to the simulation results at different stages of 
this paper. 
\begin{figure}
\begin{center}
\includegraphics[angle=-90, width=0.44\textwidth]{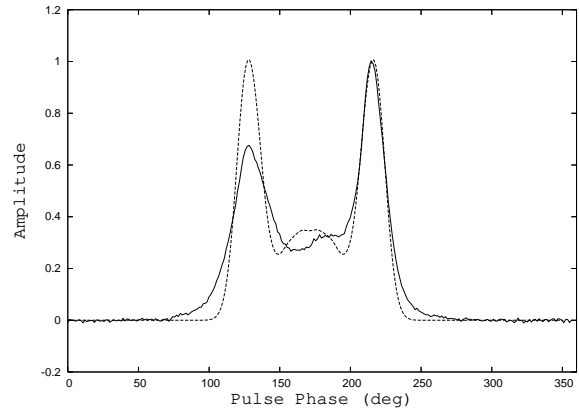}
\caption{Observed and simulated pulse profile at 325 MHz with simple dipolar geometry for the case: $\alpha=175.4$\degr, $\beta=-6.9$\degr; drift rate$D=19.05\degr/P_1$.}
\label{fig:prof_sim}
\end{center}
\end{figure}
\section{Subpulse drifting at multiple frequencies}                            \label{sec:results_drifting}
\begin{center}
\begin{figure}
\begin{center}
\includegraphics[angle=0,width=0.5\textwidth]{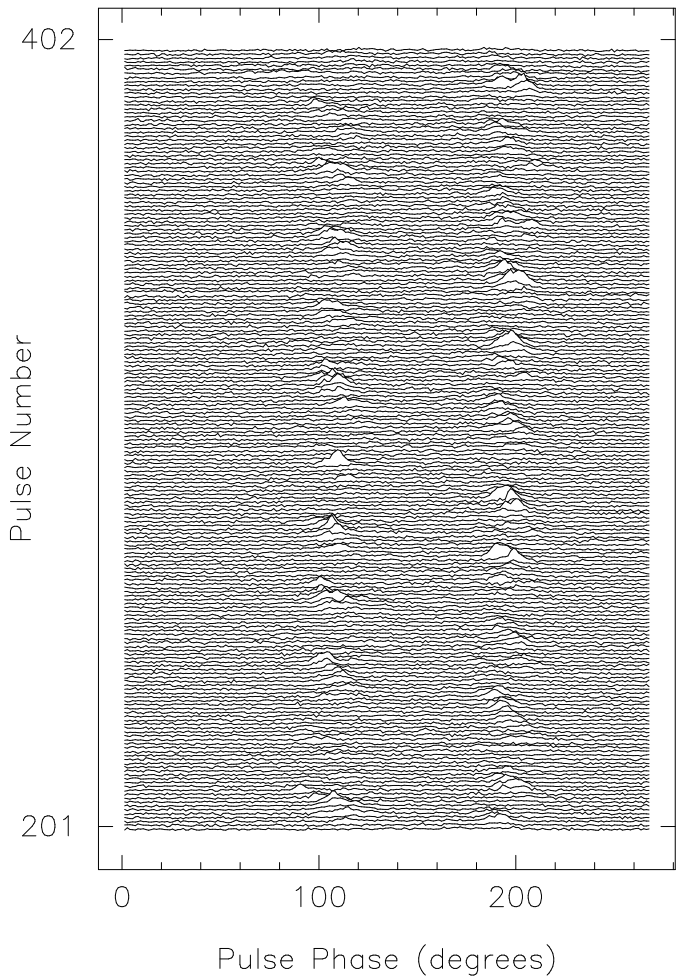}
\caption{Single pulse sequence of 200 pulses (pulse \# 200 to 400) of PSR B0818$-$41 from the GMRT observations at 325 MHz. Pulse emission is present for a wide longitude range ($\sim$ 150 \degr). Drift bands in the outer leading and trailing side are observed from the stack of single pulses. Modulations in the inner region are weaker and are indistinguishable in this plot in contrast to the outer regions.}
\label{stack_325}
\end{center}
\end{figure}
\end{center}
\begin{center}
\begin{figure}
\begin{center}
\includegraphics[angle=0,width=0.5\textwidth]{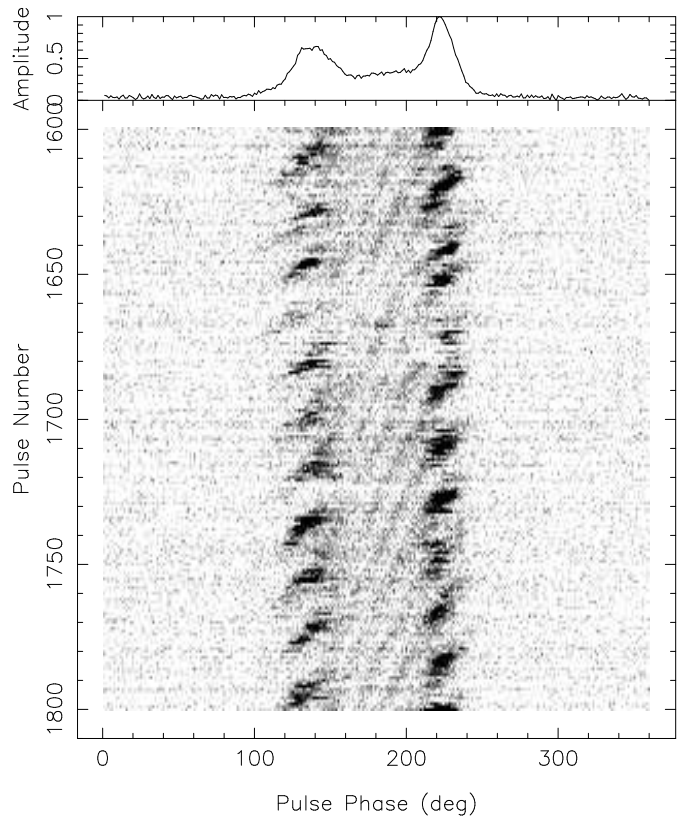}
\caption{Gray scale plot of single pulse data of PSR B0818$-$41 at 325 MHz, with the average profile shown on top. Fading of the outer left region near pulse number range 1655 to 1675, corresponding fading in the outer right region might be near pulse \# 1670.}
\label{sp_325}
\end{center}
\end{figure}
\end{center}
\begin{center}
\begin{figure}
\begin{center}
\includegraphics[angle=0,width=0.5\textwidth]{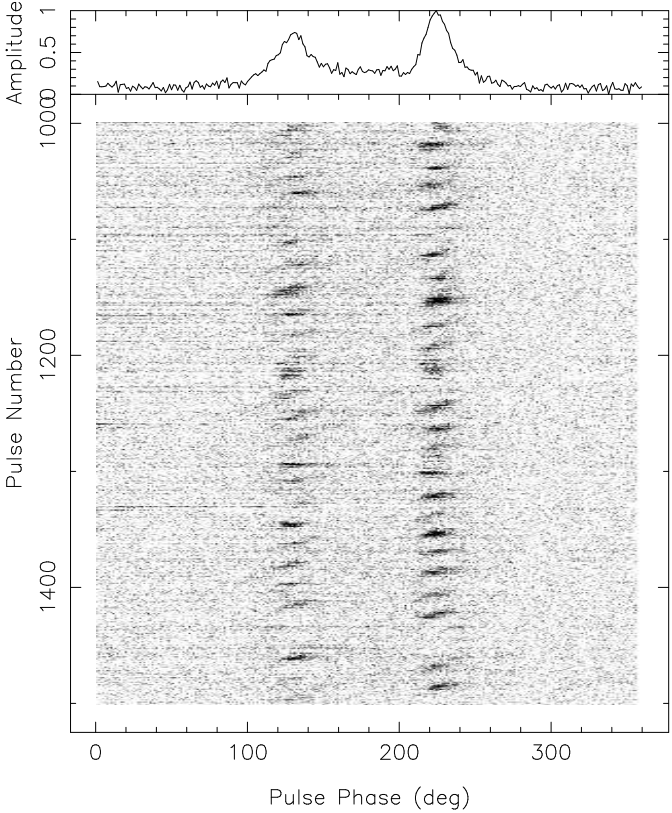}
\caption{Gray scale plot of single pulse data of PSR B0818$-$41 at 244 MHz. Drifting is observed in the leading outer and trailing outer regions, but not in
the inner region. Occasional fading of drift bands in the outer left region is seen.}
\label{sp_244}
\end{center}
\end{figure}
\end{center}
\begin{center}
\begin{figure}
\begin{center}
\includegraphics[angle=0,width=0.5\textwidth]{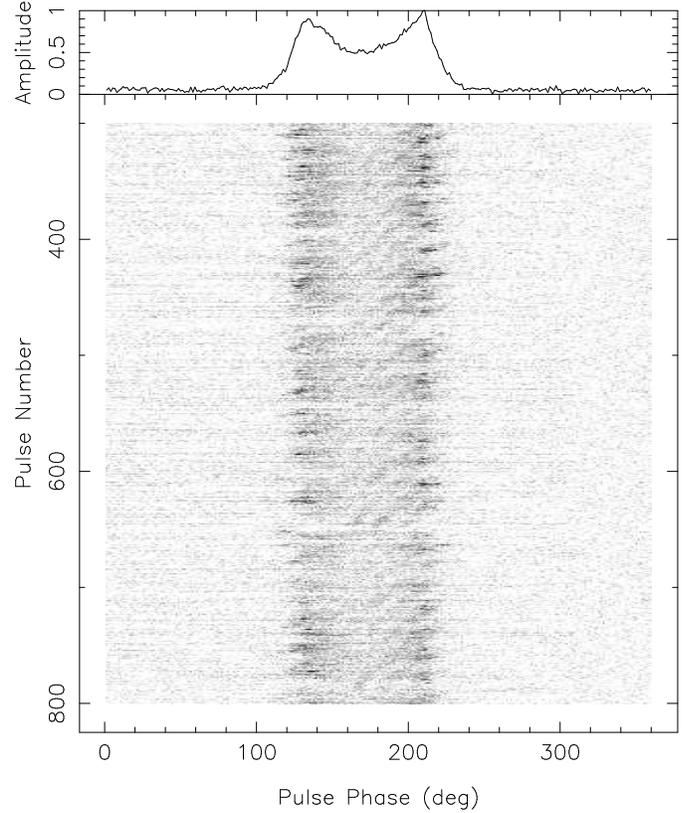}
\caption{Gray scale plot of single pulse data of PSR B0818$-$41 at 610 MHz showing regular drifting. Drift bands are observed in the inner and the outer regions.}
\label{sp_610}
\end{center}
\end{figure}
\end{center}
\begin{center}
\begin{figure}
\begin{center}
\includegraphics[angle=0,width=0.5\textwidth]{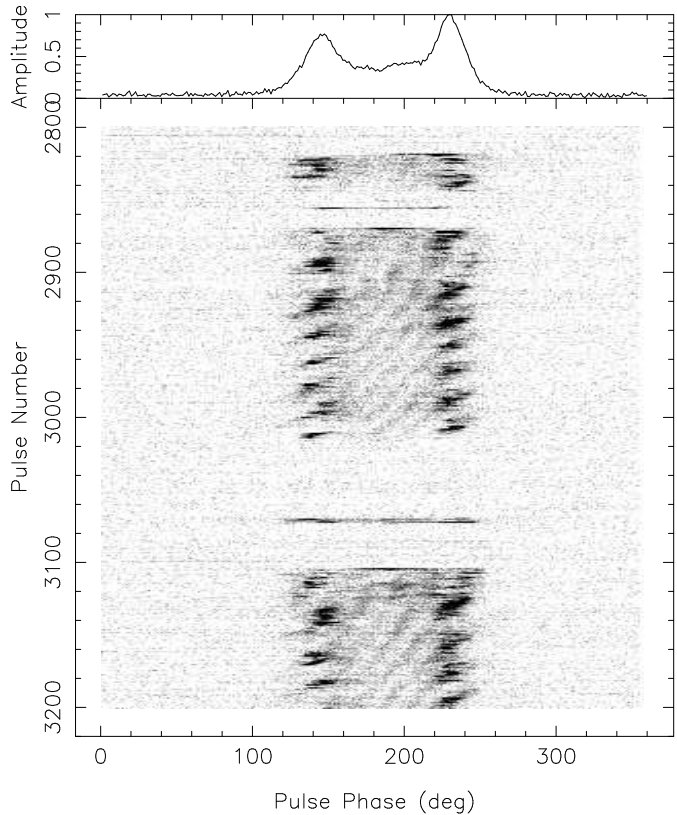}
\caption{Same as Fig.\ref{sp_325} for pulse \# 2800 to 3200. Almost stationary drift is observed around pulse \# 2870 to 2920. Then the regular negative drifting starts. Frequent nulling is observed.}
\label{sp1_325}
\end{center}
\end{figure}
\end{center}
\begin{center}
\begin{figure}
\begin{center}
\includegraphics[angle=0,width=0.5\textwidth]{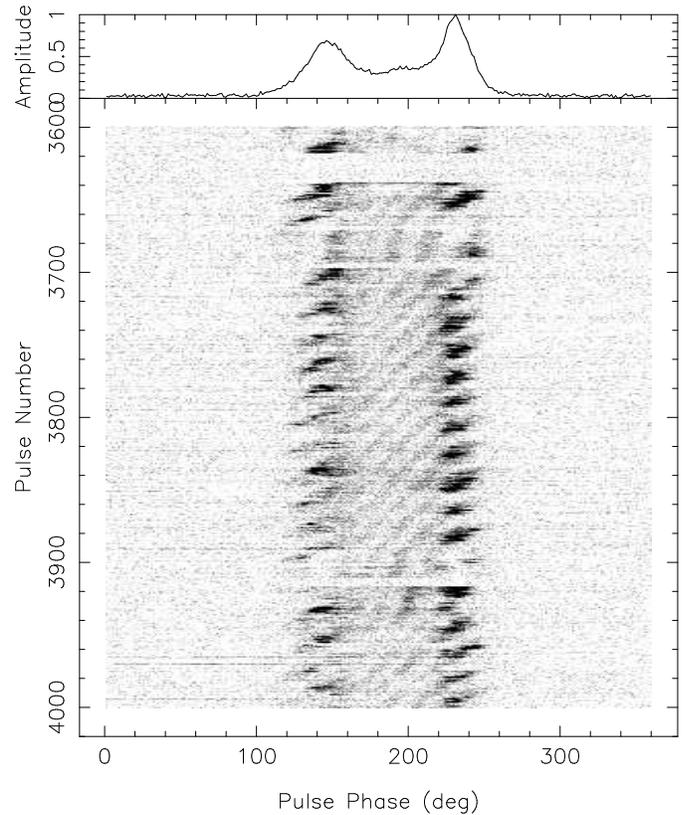}
\caption{Same as Fig.\ref{sp_325} for pulse \# 3600 to 4000. Transition from negative to stationary drift rate around pulse \# 3665, almost stationary drifting from pulse \# 3665 to 3698, regular drift starts from pulse number 3702. Drift rate again slows down around pulse \# 3900, stationary drifting is observed in following few pulses and then the pulsar starts nulling around 3913. Pulsar becomes active at around pulse \# 3920.}
\label{sp2_325}
\end{center}
\end{figure}
\end{center}
\begin{center}
\begin{figure}
\begin{center}
\includegraphics[angle=0,width=0.5\textwidth]{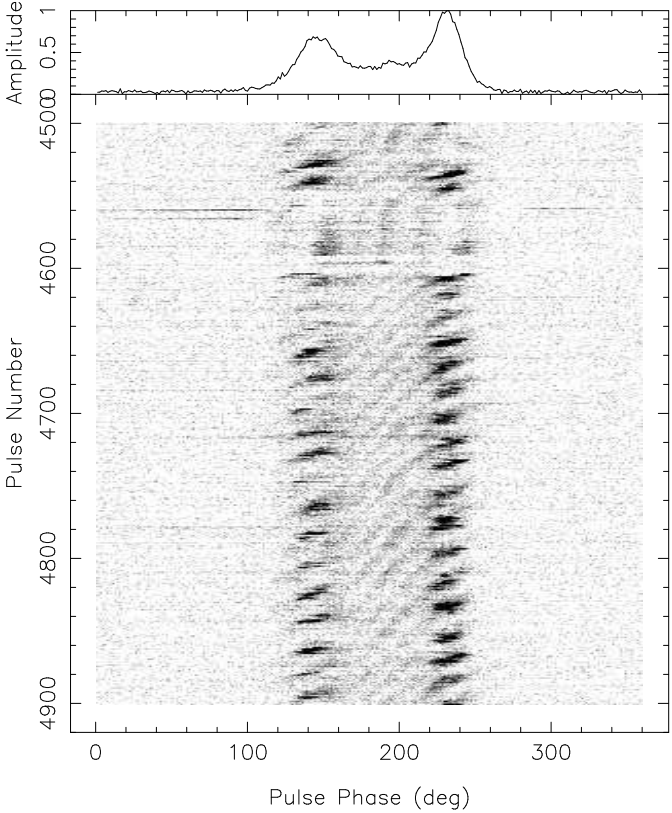}
\caption{Same as Fig.\ref{sp_325} for pulse \# 4500 to 4900. Transition from negative to stationary drift rate is observed around pulse \# 4545. Stationary drifting is observed from pulse \# 4545 to 4595, after which the pulsar starts nulling and becomes active again at pulse \# 4603.}
\label{sp3_325}
\end{center}
\end{figure}
\end{center}
\begin{center}
\begin{figure}
\begin{center}
\includegraphics[angle=0,width=0.5\textwidth]{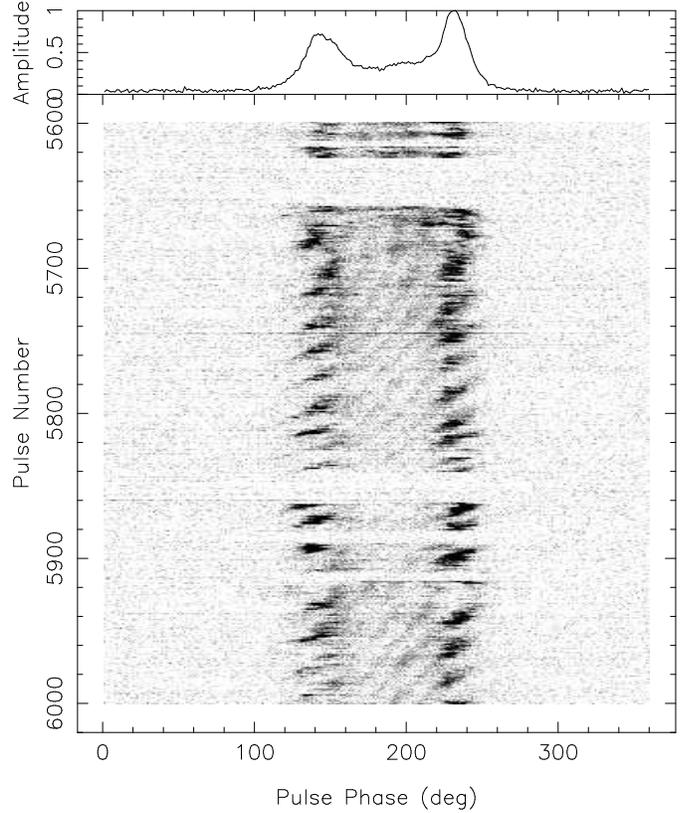}
\caption{Same as Fig.\ref{sp_325} for pulse \# 5600 to 6000. There is some indication of curved drift band around pulse \# 5660 to 5720.}
\label{sp4_325}
\end{center}
\end{figure}
\end{center}
Fig. \ref{stack_325} shows an example of single pulse data of 200 pulses at 325 MHz. Individual pulses are seen over a 
wide longitude in the pulse window. Organized drifting near the two edges of the pulse profile is observed, in the form of 
linear drift bands going from the later to the earlier longitude, as the pulse number increases (we refer to this as 
negative drifting). Fig. \ref{sp_325} is the gray scale plot of a sequence of 200 pulses showing regular drifting, at 
325 MHz. As reported in Paper 1, three different drift regions with different drift rates $-$ an inner region with 
steeper apparent drift rate flanked on each side by a region of slower apparent drift rate. Multiple drift bands (typically 
3 to 4) can be seen in the inner region for any single pulse, whereas the outer regions show only a single drift band. 
Furthermore, the two drift regions maintain a constant phase relationship. Though, simultaneous multiple drift regions 
are more prominent at 325 MHz, drifting is also observed at 150, 244 and 610 MHz. Fig.\ref{sp_610} shows the gray scale 
plot of the single pulses at 610 MHz. Though the single pulses become quite weak at 610 MHz, drifting is clearly visible in 
the gray scale plot both in the inner and the outer region. In the inner region, 3 to 4 simultaneous drift bands are visible and 
faint single drift bands can be identified on both sides of the inner region. Fig. \ref{sp_244} is the gray scale plot of 
single pulses at 244 MHz. Drifting is observed in the leading and trailing outer regions, but no signature of drifting is 
present in the inner saddle region. Even after integrating the data up to a resolution of 8 ms to detect possible weak drift 
bands in the inner drift region, we do not see any evidence of that. At 150 MHz faint drift bands are observed in the trailing 
outer region, but no evidence of drifting is observed either in the inner region or in the leading outer region. 

Occasionally, either leading or trailing outer region partially or totally disappears, while normal emission is observed 
from the other regions. Fading of the leading outer region is more commonly observed, e.g. around pulse \# 1655 to 1675 in 
Fig. \ref{sp_325}; 1420 to 1440 in Fig. \ref{sp_244} and 470 to 490 in Fig. \ref{sp_610}. 

We also observe frequent nulling in this pulsar. Fig. \ref{sp1_325} shows some sequences of pulsar nulls. Duration of the 
nulls varies from a few pulses to few hundreds of pulses. Nulling property of this pulsar will be discussed in detail in a 
follow up paper.\\

\subsection{Frequent change of drift rates :}
Though the regular drift patterns observed in Fig. \ref{sp_325} are quite common for this pulsar, we observe frequent changes 
of the drift rates. Some extreme examples of changing drift rates such as transitions from negative to stationary or stationary 
to negative drifting are also observed. Because of the observability of simultaneous multiple drift regions with good signal to 
noise, changing drift rates are more prominently seen at 325 MHz. Due to the absence of drifting in the inner region at 244 MHz and 
weaker drift bands at 610 MHz, such regions are harder to identify at these frequencies. We investigate changes in drift rates 
for about 10,000 pulses from two different epochs of observations at 325 MHz. In addition to frequent occurrences of small changes 
in the drift rate, we could find, seven transitions from the negative to stationary drift rates, five transitions from stationary 
to negative drift rates, and two possible signatures of curved drift bands. We also observe one pulse sequence showing stationary 
drifting interleaved between nulling regions. Fig. \ref{sp1_325}, \ref{sp2_325} and  \ref{sp3_325} illustrate some of these 
phenomenon. Fig. \ref{sp1_325} shows the single pulse gray scale plot for pulse \# 2800 to 3200. 
Almost stationary drift is observed around pulse \# 2870 to 2920 after which the regular negative drifting starts. Irregular 
drifting is observed again around pulse \# 3107. Fig. \ref{sp2_325} presents the single pulse gray scale plot for pulse \# 
3600 to 4000. Stationary drift bands are observed from pulse \# 3665 to 3698 after which regular negative drifting starts from 
pulse \# 3702. The apparent drift rate again slows down around pulse \# 3900, stationary drifting is observed in following few pulses 
and then the pulsar starts nulling around 3913. Fig. \ref{sp3_325} presents the single pulse gray scale plot for pulse \# 4500 
to 4900. Transition from negative to stationary drift rate is observed around pulse \# 4545. Stationary drifting is observed from 
pulse \# 4545 to 4595, after which the pulsar starts nulling and becomes active again at pulse \# 4603. Fig. \ref{sp4_325} presents 
the single pulse gray scale plot for pulse \# 5600 to 6000. Indication of curved drift bands is seen around pulse \# 5660 to 5720.
For the sequences showing transitions from negative to stationary drift, nulling starts immediately after the stationary drifting. 
Similarly for the sequences showing transitions of stationary to negative drift, the stationary drifting follows just after nulling. 
Almost no situation is seen (apart from some indication in curved drift bands) where the drift direction is positive. Possible 
negative-stationary-negative or negative-stationary-positive transitions are generally interrupted by nulls. 

To summarise, frequent changes of drift rates $-$ including transitions from negative to stationary and stationary to negative drift 
rates $-$ are observed for PSR B0818$-$41. We observe transitions from negative to stationary drift or stationary to the negative 
drift in different pulse sequences at 325 MHz. Such transitions of negative to stationary drift are always followed by nulling. 
After nulls the drifting is stationary (or irregular) for few pulse and then the transition from stationary to negative drift 
is observed.

{\bf Interpretations with aliasing:}
According to \cite{Ruderman_etal} model, observed stationary drifting for unaliased case will imply that $E\times B=0$, caused by $E=0$, 
and hence no radiation. Consequently it will be difficult to explain the observed stationary drifting and the transitions from negative 
to stationary drift rates without aliasing. In the following we try to explain the observed drifting with aliasing.
 
According to \cite{Ruderman}, for the case of unaliased drifting one should observe negative drifting for inner LOS geometry and positive 
drifting for outer LOS geometry. Although, for {\bf G-1} the observed negative drift, can be explained either by unaliased drifting or by 
aliased drifting with even alias order (Paper 1), observed occasional changes of drift rates can be easily understood with aliased drifting. 
Observed negative drift is explained if, $0<P_3^t<P_1$, and the transitions from negative to stationary drift rates can be explained by 
slight decrease of the drift rate, so that $P_3^t$ is equal to $P_1$. The intrinsic drift rate for the unaliased case is positive for 
{\bf G-2}. Thus to reproduce the observed negative drifting for {\bf G-2}, we need to consider aliased drifting so that $P_1<P_3^t<2P_1$ 
and the apparent drift is of opposite sign than that of the true drift rate i.e. the true positive drift rate will appear as negative drift 
rate to the observer. Occasionally drift rate slightly increases so that $P_3^t$ is equal to $P_1$, and we observe stationary drift bands. 
Hence for {\bf G-1}, $P_3^t$ is slightly less than $P_1$ will give rise to aliased negative drifting and transition from the negative to 
the stationary drifting means slowing down of the carousel so that $P_3^t=P_1$. For {\bf G-2}, $P_3^t$ is slightly more than $P_1$ will give 
rise to aliased negative drifting and transition from the negative to the stationary drifting means speeding up of the carousel. Similarly 
the observed stationary to negative drift rate transitions can be interpreted as speeding up of carousel for {\bf G-1} and slowing down of 
the carousel for {\bf G-2}. 

Frequent occurrence of nulling either before or after of stationary drifting indicates that somehow nulling is associated with the condition
$P_3^t=P_1$. After transitions of apparent drift rate from negative-stationary direction, we generally observe nulling, implying that, pulsar
stops emitting after carousel slows down for {\bf G-1} and after the carousel speeds up for {\bf G-2} to satisfy $P_3^t=P_1$. Quite often, 
recovery of pulsar emission after null, starts from stationary drifting ($P_3^t=P_1$) and then the regular negative drifting follows. 
Association of stationary drifting and nulling indicates that the mechanism responsible for pulsar drifting is closely associated with the 
pulsar emission mechanism.

Sign reversals of drifting and curved drift bands are observed for PSR B0826$-$34 \citep{Biggs_etal}, unlike PSR B0818$-$41 where we do not
observe positive drifting (except for some indications of curved drift bands interrupted by nulling, e.g. Fig. \ref{sp4_325}).
Similar to our interpretations for explaining the observed transitions of negative to stationary drift rates in B0818$-$41, \cite{Gupta_etal}
proposed that the observed drift rate for PSR B0826$-$34 is aliased version of the true drift rate which is such that a subpulse drifts to the 
location of the adjacent subpulse (or a multiple thereof) in about one pulse period. They showed that small variations in the mean drift rate 
are enough to explain the apparent reversal of drift directions seen in the data. 

\cite{Gupta_etal} proposed that the small variations in the mean drift rate caused by very small heating and cooling effects of the 
polar cap can explain the apparent reversals of drift directions observed for PSR B0826$-$34. Similarly, the slight change of the drift 
rate required for transition form negative to stationary directions of apparent drifting for PSR B0818$-$41 can be explained by very small 
heating and cooling effects in the surface temperature of the polar cap \citep{Gil_etal_03}. We note that the transitions of drift rates are 
rather abrupt for quite a few sequences. Which can be interpreted by means of quantum jump or quite fast change in polar cap temperature, 
causing $P_3^t$ slightly less than $P_1$ (giving rise to observed negative drift) to $P_3^t=P_1$ (giving rise to stationary drifting) for 
{\bf G-1}, and vice verse.

\subsection {Determination of $P_{3}^{m}$ :} 
\begin{center}
\begin{figure}
\begin{center}
\includegraphics[angle=-90,width=0.4\textwidth]{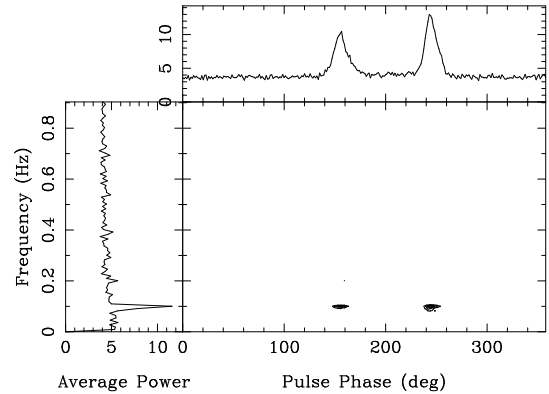}
\caption{The contour plot of the power spectrum of the flux at 325 MHz as a function of pulse phase for the sequence of 200 pulses showing regular drifting as in Fig.\ref{sp_325}. The left panel shows the power spectrum integrated over the entire pulse longitude. The upper panel shows the power integrated over fluctuation frequency. The peak fluctuation feature is at 18.3 $P_1$.}
\label{flusp_325}
\end{center}
\end{figure}
\end{center}

For the determination of $P_{3}^{m}$, we use the fluctuation spectrum analysis technique \citep{Backer_70}. Fig. \ref{flusp_325} plots the 
phase resolved fluctuation spectrum for a sequence of 200 pulses showing regular drifting in Fig.\ref{sp_325}. The strong peak at 18.3 $P_1$ 
is identified as $P_3^m$. The value of $P_3^m$ varies slightly for different pulse sequences with regular drifting, from minimum of about 
16.7 $P_1$ to maximum of 21.4 $P_1$. For a given pulse sequence, $P_3^m$ appears to be the same for leading outer region, inner saddle region and 
trailing outer region. This is true even for the pulse sequences showing irregular drifting. 

\begin{center}
\begin{figure}
\begin{center}
\includegraphics[angle=-90,width=0.4\textwidth]{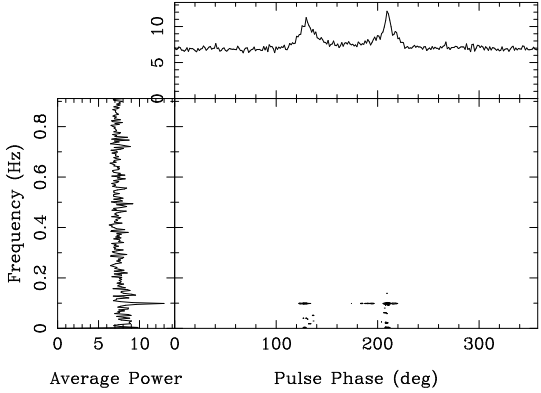}
\caption{Same as Fig.\ref{flusp_325}, but at 610 MHz for the sequence of 500 pulses in Fig.\ref{sp_610}. The peak fluctuation feature is at around 18.6 $P_1$.}
\label{flusp_610}
\end{center}
\end{figure}
\end{center}

Fig. \ref{flusp_610} shows the fluctuation spectrum for the sequence of 800 pulses with regular drifting at 610 MHz shown in Fig. \ref{sp_610}. 
We observe a strong peak corresponding to $P_3^m$ at 18.6 $P_1$, which is the same for both the inner and the outer regions. 

\begin{center}
\begin{figure}
\begin{center}
\includegraphics[angle=-90,width=0.4\textwidth]{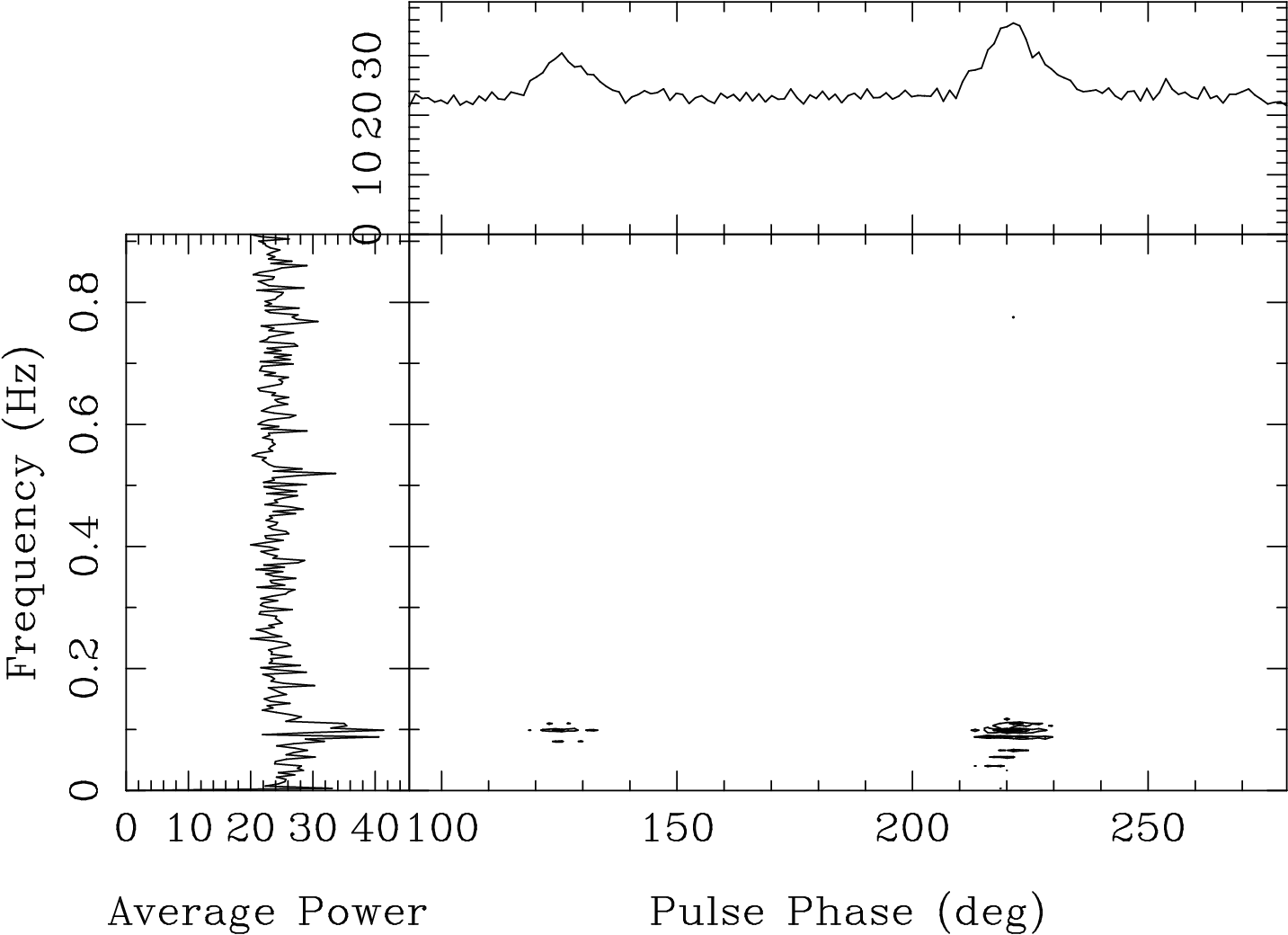}
\caption{Same as Fig.\ref{flusp_325}, but at 244 MHz for the sequence of 500 pulses in Fig.\ref{sp_244}. The peak fluctuation feature is at 18.6 $P_1$.}
\label{flusp_244}
\end{center}
\end{figure}
\end{center}

Fig. \ref{flusp_244} shows the fluctuation spectrum at 244 MHz for the sequence of 500 pulses shown in Fig. \ref{sp_244}. We observe a strong 
peak corresponding to $P_3^m$ at 18.6 $P_1$, for the leading outer and the trailing outer regions. Even after integrating the data up to 8 ms, 
the fluctuation spectrum analysis does not show any periodicity present in the inner region. 

Table \ref{table2} lists the $P_3^m$ values, at the individual frequencies, calculated for pulse sequences with regular drifting. Within error bars,
 the value of $P_3^m$ ($\sim 18 P_1$) is the same at all the frequencies. This supports the general result arrived at, for a larger sample of 
pulsars by \cite{Weltevrede_etal_06} and (2007) at 20 and 92 cm respectively. 
\subsection{Determination of $P_{2}^{m}$ :}             \label{sec:P2_det}
From the gray scale plot of the single pulses at 325 MHz (e.g. Fig.\ref{sp_325}), it is evident that the $P_{2}^{m}$ 
values for the outer regions are more than that of the inner region. Also it appears that the drift bands in the 
trailing outer region are somewhat steeper than the drift bands in the leading outer region. $P_{2}^{m}$ value in the inner 
drift region at 325 MHz is calculated from the auto correlation function of the single pulses averaged over the number 
of available pulses. The secondary peak of the correlation function will correspond to the correlation between adjacent 
drift bands and will give us the $P_{2}^{m}$ value. From the location of this secondary peak, $P_{2}^{m}$ value for the 
inner region is found to be 17.5$\pm$1.3\degr for 200 pulses (pulse \# 1600 to 1800) showing regular drifting (Fig.\ref{sp_325}). 
We determine similar $P_{2}^{m}$ values, when calculated for another pulse sequence of 200 pulses showing regular 
drifting as well as for the full data set containing all the 6600 pulses. Similar correlation analysis performed for 
the inner region at 244 MHz, even with 8 ms integration, does not show any secondary peak. Due to poor signal to noise, 
the secondary peak position at 610 MHz can not be clearly determined from zero pulse offset correlation. From the one 
pulse offset correlation (i.e. while computing the auto correlation function, we correlated each pulse to the adjacent 
pulse, as done in \cite{Bhattacharyya_etal}), we determine, $P_{2}^{m}$=16.5$\pm$1.3\degr for the inner drift region at 610 MHz. 
\begin{center}
\begin{figure}
\begin{center}
\includegraphics[angle=-90,width=0.4\textwidth]{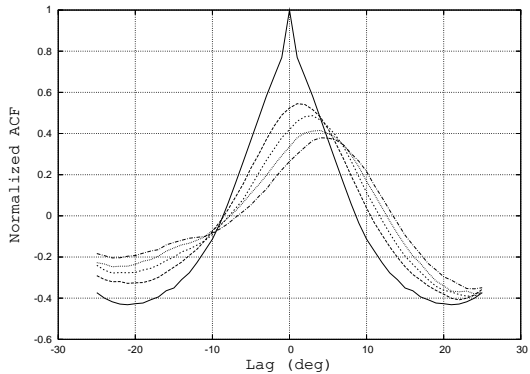}
\caption{Auto correlation results for the leading outer region at 325 MHz. Curves are for 4 different pulse offsets :
solid curve (pulse offset=0), long dash (pulse offset=$-$1), short dash (pulse offset=$-$2), dotted (pulse offset=$-$3),
dash-dot (pulse offset=$-$4).}
\label{shift_cor}
\end{center}
\end{figure}
\end{center}
\begin{center}
\begin{figure}
\begin{center}
\includegraphics[angle=-90,width=0.4\textwidth]{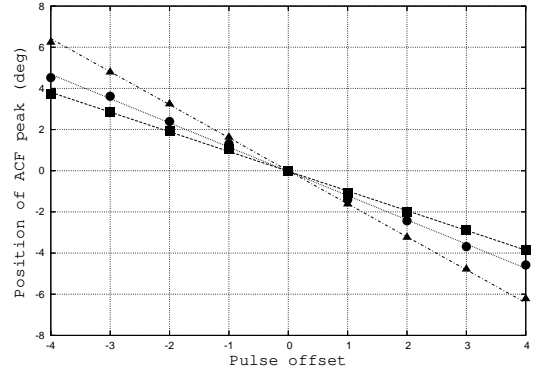}
\caption{Plot of position of the peak of the correlation functions versus the corresponding pulse offsets. 
Squares and dashed line : for the inner region and the corresponding fitted straight line with slope
$-0.96$\degr/$P_1$; circles and dotted line : for the trailing outer region and the corresponding fitted 
straight line with slope $-1.2$\degr/$P_1$; triangle and dash-dot line : for the leading outer region and 
the corresponding fitted straight line with slope $-1.7$\degr/$P_1$}
\label{slope_p2}
\end{center}
\end{figure}
\end{center}

Since simultaneous multiple drift bands are not visible in the outer regions, $P_2^m$ values can not be estimated using 
the method described above. Instead we use the following technique to determine the $P_2^m$ value of the outer regions. The 
correlation function, as described above, is calculated for different pulse offsets. The peak of the correlation function 
with zero pulse offset is at zero lag and with increase in pulse offset the peak shifts to higher lags in pulse longitude 
(e.g. Fig. \ref{shift_cor}). We plot the position of the peak of the correlation function versus the pulse offset. This 
plot is a straight line whose slope should reproduce the slope of a drift band (e.g. Fig. \ref{slope_p2}). From the slope 
of the fitted straight line and the $P_3^m$ values, one can estimate the $P_{2}^{m}$. We refer to this method as pulse 
offset correlation method (POCM hereafter). A pulse sequence showing regular drifting (pulse \# 1600 to 1800 shown in 
Fig.\ref{sp_325}), is considered for determination of $P_{2}^{m}$ values. As a crosscheck, we apply POCM to determine 
the $P_{2}^{m}$ value of the inner region. The $P_2^m$ values for the inner drift region at 325 and 610 MHz determined 
using POCM are identical to the respective values determined from the secondary peaks of the correlation function.   
Fig. \ref{shift_cor} is the plot of the auto correlation functions of the leading outer region calculated for 
different pulse offsets at 325 MHz. The peak of the correlation function shifts with increasing pulse offsets. The 
positions of the peaks of the correlation function versus the pulse offset are plotted for the leading outer, inner 
and trailing outer drift regions in Fig. \ref{slope_p2}. From the slopes of the straight lines fitted to those, we determine 
$P_2^m$=21.6\degr for the trailing outer region and $P_2^m$=31\degr for the leading outer region. Determined $P_2^m$ values 
for leading outer, inner and trailing outer regions calculated for another sequence of 200 pulses showing regular drifting, 
are also in the same ballpark. We apply this method for determination $P_2^m$ at 244 and 610 MHz, and obtain, $P_2^m$=16.9\degr 
for the leading outer, $P_2^m$=29.8\degr for the trailing outer regions at 244 MHz and  $P_2^m$=16.5\degr for the inner 
region at 610 MHz. But surprisingly, when calculated for different pulse offsets for the outer regions at 610 
MHz, very less shift (for trailing outer region) or almost no shift (for leading outer region) of the peak of 
the correlation function is observed which may imply that the longitudinal extents of the outer drift region 
is quite small at 610 MHz. So we were unable to determine $P_2^m$ values for the outer regions 
at 610 MHz. The $P_2^m$ values for the leading outer and the trailing outer region at different frequencies 
are listed in Table \ref{table2}. 

The absence of the inner drift region at 244 MHz can be interpreted as, the LOS missing the emission from the inner ring, which 
is supported by the results from the simulations (Sect. \ref{sec:sim}).
$P_{2}^{m}$ value, for the inner drift region at 325 MHz, determined in this work is close to the value estimated 
in Paper 1. $P_2^m$ in the outer regions are more than that of the inner region, which is also evident from the 
visual inspection of the drift pattern at 325 MHz. The $P_2^m$ for the leading outer drift region is more than 
the $P_2^m$ for the trailing outer region, both at 244 and 325 MHz (Table \ref{table2}). This is difficult to explain 
with a concept of circular outer ring around the magnetic pole with isotropic distribution of sparks. With the 
simulations (Sect. \ref{sec:sim}), we obtain identical $P_2^m$ values for the leading and trailing outer regions as 
well as for the inner region. Spark patterns centered around a local pole rather than the magnetic dipole axis may 
provide a possible explanation.
\subsection {Determination of the subpulse width ($\Delta\Phi_s$)}                    \label{sec:results_subpulsewidth}
To determine the subpulse width ($\Delta\Phi_s$), we compute the autocorrelation function, with zero pulse offset, and find 
the full width at half maximum (FWHM). Values of $\Delta\Phi_s$, calculated for a pulse sequence with 200 pulses showing regular 
drifting at 325 MHz, are equal to 12.3\degr, 6.3\degr and 10.8\degr for the leading outer, inner and trailing outer regions 
respectively. $\Delta\Phi_s$ values calculated for another pulse sequence with 200 pulses showing regular drifting, as well as for the 
full data set containing 6600 pulses at 325 MHz, are identical. We calculate the $\Delta\Phi_s$ values for the inner and the outer 
regions at 610 MHz and for the outer regions at 244 MHz. Table \ref{table2} lists the values of $\Delta\Phi_s$ for all the three drift 
regions determined from our observations at different frequencies. For both 325 and 610 MHz, $\Delta\Phi_s$ values in the inner 
region are less than that of the outer regions. $\Delta\Phi_s$ values in leading outer region are more than that of trailing outer 
region at 325 and 610 MHz. However, at 244 MHz, $\Delta\Phi_s$ values for the leading outer and trailing outer regions are comparable 
and more than the corresponding values at 325 MHz. Evolution of $\Delta\Phi_s$ values with pulse longitude follows the trend of evolution 
of $P_2^m$. Though within error bars, $\Delta\Phi_s$ value, for the inner drift region at 610 MHz is less than the corresponding value 
at 325 MHz, which is expected by the radius to frequency mapping. Hence, evolution of $\Delta\Phi_s$ and $P_2^m$ values with frequency 
and longitude mostly follow similar trend.
\begin{center}
\begin{figure}
\begin{center}
\includegraphics[angle=0,width=0.4\textwidth]{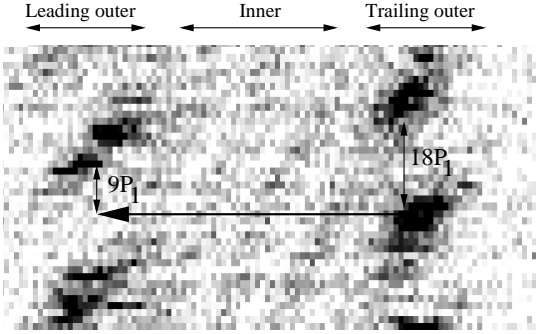}
\caption{Gray scale plot of 27 single pulses from PSR B0818$-$41 at 325 MHz. The horizontal arrow is drawn to illustrate that the peak emission from trailing outer and leading outer region are not in phase (offset by $\sim$ 9 $P_1$).}
\label{sp_zoom}
\end{center}
\end{figure}
\end{center}
\begin{center}
\begin{figure}
\begin{center}
\includegraphics[angle=-90,width=0.4\textwidth]{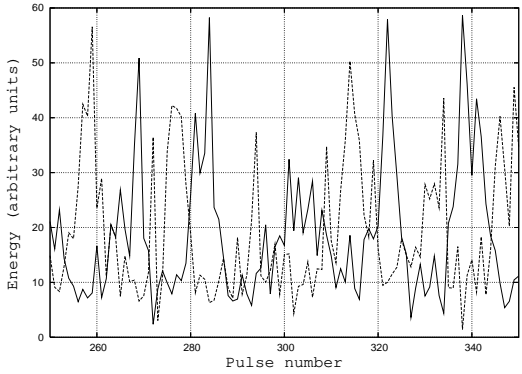}
\caption{Pulse energy under the trailing outer region (solid line) and the leading outer region (dashed line) for pulse \# 250$-$350 at 325 MHz.}
\label{pe_325}
\end{center}
\end{figure}
\end{center}
\section {Phase relation between multiple drift regions}                                 \label{sec:results_pecor}
Simultaneous multiple drift regions maintain unique phase relationship for PSR B0818$-$41 : (1) peaks of the 
emission from outer leading and trailing regions are at constant phase offset (observed at 244, 325 and 610 MHz) and 
(2) emission from the inner and outer regions are always locked in phase (observed at 325 and 610 MHz). These phase 
relationships are clearly observed in Fig. \ref{sp_zoom}, a zoomed  plot of 27 single pulses at 325 MHz. 

\subsection {Investigation of the constant phase offset between the peaks of emission from leading and trailing outer regions}
\begin{center}
\begin{figure}
\begin{center}
\includegraphics[angle=-90,width=0.4\textwidth]{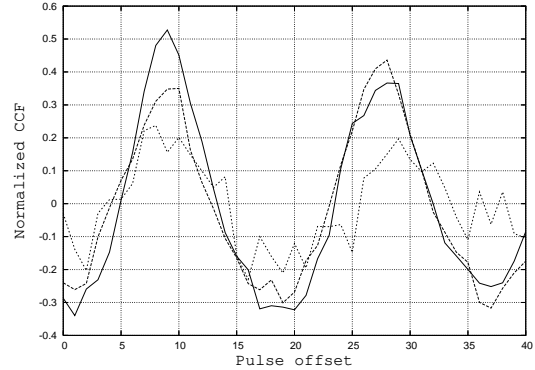}
\caption{Cross correlation of pulse energy in the leading and trailing outer regions as a function of pulse offset, calculated for pulse sequences showing regular drifting. Solid line (325 MHz), long-dash (610 MHz), short-dash (244 MHz).}
\label{cross_cor}
\end{center}
\end{figure}
\end{center}

As seen in the single pulse gray scale plots (e.g. Fig. \ref{sp_zoom}), the peaks of emission from the leading and 
trailing outer regions as a function of pulse number are not in phase and are offset by a constant pulse period, which 
we denote as, $P_5$$\sim$ 9$P_1$ from eye estimation. We use the following procedure to study this phase relation more 
quantitatively. For each pulse we compute the total energy under the trailing and the leading peaks (up to a point 
where the energy goes down to 50\% of the peak value on both sides of the peak). Fig. \ref{pe_325} shows the variation 
of the energy of the leading and the trailing peaks with pulse number, for a sequence of 100 pulses showing regular drifting. 
Peaks of the emission for leading and trailing outer regions are observed to be offset by $P_5$ $\sim$ 9$P_1$. To get a 
more accurate estimation of $P_5$, we cross correlate the energy under the trailing and leading peaks for different pulse lags. 
A similar exercise is performed at 244 and 610 MHz. Fig. \ref{cross_cor} presents the plot of the correlation between the energy 
of trailing and leading outer regions as a function of pulse lag, for 244, 325 and 610 MHz. Initially, the emissions from 
leading and trailing peaks are anticorrelated; the correlation builds up with increasing pulse offsets and reaches to maximum 
value near $\sim$ 9$P_1$, after which the secondary peaks are observed at 18$P_1$ interval. To avoid contributions from 
frequent nulls and irregular drifting, this exercise was performed on sequences of pulses showing regular drifting (e.g. 
Fig. \ref{sp_325}). We repeat this exercise for 13 such pulse sequences, each containing about 200 pulses showing regular 
drifting. $P_5$ varies slightly, from minimum of 6$P_1$ to maximum of 10$P_1$, with an average of 9.2$P_1$. No apparent 
frequency dependence of $P_5$ is observed.  
{\bf Solution of the aliasing problem :}
Drifting subpulses are an excellent example of under sampling. The aliasing problem is solved for only a few drifting 
pulsars (e.g. PSR B0809$+$74, B0943$+$10 etc) using extra information from specific features present in the data. In the 
following we describe a scheme for solving the aliasing problem using knowledge of $P_5$. 

\begin{center}
\begin{figure}
\begin{center}
\includegraphics[angle=0,width=0.5\textwidth]{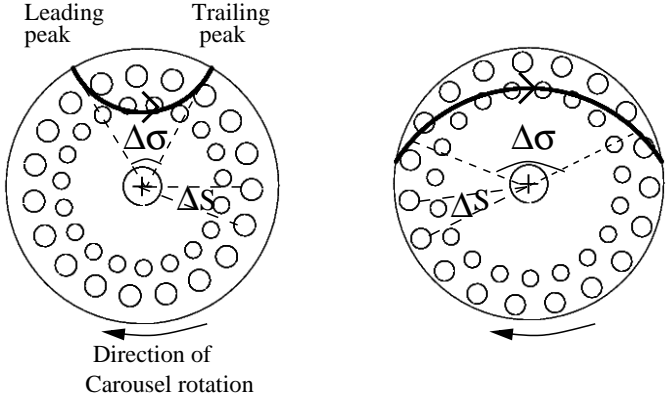}
\caption{Arrangements of sparks in two rings in the polar cap and possible LOS cuts for {\bf G-1} (left panel) and {\bf G-2} (right panel).}
\label{polcap}
\end{center}
\end{figure}
\end{center}

\begin{figure}
\begin{center}
\includegraphics[angle=-90, width=0.44\textwidth]{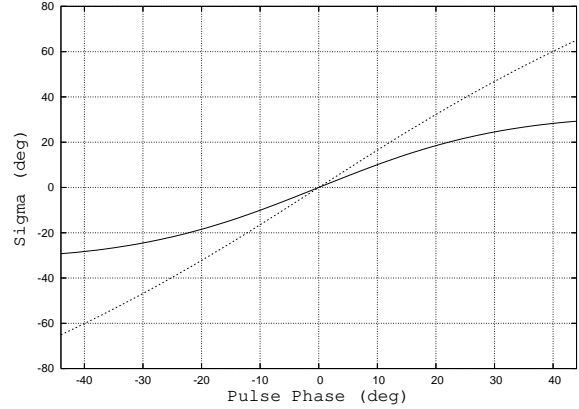}
\caption{Variation of $\sigma$ with $\phi$ (see Eqn.\ref{eqn1} of text) for {\bf G-1} 
($\alpha=11$\degr, $\beta=-5.4$\degr, solid curve) and {\bf G-2} ($\alpha=175.4$\degr, $\beta=-6.9$\degr, dashed curve).}
\label{fig:eta_var}
\end{center}
\end{figure}

Fig. \ref{polcap} illustrates the configuration of sparks in two concentric polar rings and the possible LOS cuts, for both 
{\bf G-1} and {\bf G-2}. Here $\Delta \sigma$ is the profile width measured with respect to magnetic axis and 
$\Delta s\:=\:360/N_{sp}$, is the angular separation between two consecutive sparks measured around the magnetic axis. 
Now we consider the ratio,

\begin{equation}  
\frac{\Delta\sigma}{\Delta s}=I_1+f_1
\label{eqn2}
\end{equation}
where $I_1$ and $f_1$ denote the integer and fractional parts of the ratio. This gives the total number of sparks occupying 
the outer ring between the points of intersection with the leading and trailing peaks of the profile. Here, $f_1$ is a measure of 
observed phase offset : $f_1=0$ will give drift bands which are in phase, whereas $f_1=0.5$ will produce a phase offset of 9$P_1$.
The above analysis assumes that the spark pattern drifts by a negligible amount in the duration of the LOS traverse from the leading 
to trailing peak. In reality, the spark pattern will drift by $\Delta\phi\:D/360$\degr during this interval, where D is the amount by 
which the sparks drift in one pulse period ($P_3^m=360/D$) and $\Delta\phi$ is the separation between two peaks of pulse profile.
\begin{equation}  
\frac{\Delta\sigma}{\Delta s}+\frac{\frac{\Delta\phi\:D}{360}}{\Delta s}=I_2+f_2
\label{eqn3}
\end{equation}
For a finite $f_2$, the peak emission from leading and trailing outer regions will be out of phase by a constant pulse offset 
($P_5$), which will be determined by the time taken by the sparks to traverse $(1-f_2)\Delta s$\degr. 
\begin{equation} 
P_5=\frac{(1-f_2)\Delta s} {D_{ap}}
\label{eqn4}
\end{equation}
where, $D_{ap}$ (apparent drift rate) is the amount by which a spark pattern with drift rate $D$ will appear to move in one pulse 
period and is given by $D_{ap}=\bmod(k\:D-\Delta s)$, $k=INT(n+1)/2$, $n$ being the alias order (Paper 1). With the knowledge of 
viewing geometry, $N_{sp}$, alias order $n$ and drift rate $D$, one can easily calculate the expected $P_5$ using Eqn. \ref{eqn4}; 
which can then be compared with the observed value. Therefore, knowing viewing geometry, $N_{sp}$ and observed value of $P_5$, it 
is possible to constrain drift rate $D$ and possible alias order. This technique can solve the aliasing problem and will provide 
an independent estimation of $P_3^t$ and $P_4$. The only missing term is an estimate for $\Delta\sigma$. This is obtained from the 
pulse width $\Delta\phi$, using the known geometry.
The general formula that relates the azimuthal angle measured with respect to the magnetic axis ($\sigma$) with the rotational phase 
($\phi$) is,
\begin{equation}  
sin(\sigma)=\frac{sin(\alpha+\beta)\:sin(\phi)}{sin(\Gamma)} 
\label{eqn1}
\end{equation}
where $\Gamma$ is the angle between the magnetic axis and the LOS at the rotational phase $\phi$ \citep{Gupta_etal}. Fig. \ref{fig:eta_var} 
shows the variation of $\sigma$ as a function of $\phi$ around the magnetic meridian ($\phi$=0 and $\sigma$=0) for {\bf G-1} (solid curve) 
and {\bf G-2} (dashed curve). For both {\bf G-1} and {\bf G-2}, $\Delta \Phi$ is equal to 88\degr at 325 MHz (Table \ref{table2}). This 
corresponds to $\Delta\sigma \sim 58$\degr for {\bf G-1} and $\Delta \sigma \sim 130$\degr for the {\bf G-2} (Fig. \ref{fig:eta_var}).
We consider two concentric rings each with 19 sparks. Using Eqn. \ref{eqn2} for {\bf G-1}, we have $\Delta\sigma/\Delta s=3.08$, 
i.e. $I_1=3$ and $f_1=0.08$. Fractional part $f_1$ is very small. Therefore, for negligible drifting, as the LOS passes from the 
first to the third spark, it traverses from the leading to the trailing peaks. In this case the peak emission from the leading 
and trailing outer regions will be in phase, which is also reproduced with simulations with {\bf G-1}. Similarly for {\bf G-2}, 
$\Delta\sigma/\Delta s=6.86$ i.e. $I_1=6$ and $f_1=0.86$. Therefore, with negligible drifting, passage of LOS from the first to the 
sixth pulse and an extra amount, $f_1 \Delta s$\degr, will imply the traverse from the leading to the trailing peak which is also 
reproduced in simulations with {\bf G-2}. 

We calculate pulse offset $P_5$ from Eqn. \ref{eqn4} for both the geometries for the cases of unaliased drifting and aliased drifting 
with different $N_{sp}$ values ranging from 17 to 22. Table \ref{table:d_p_inner} lists different combinations of $D$, $n$, $N_{sp}$ 
values required to reproduce the observed drift pattern and the corresponding estimations of $P_5$ for {\bf G-1}. Table \ref{table:d_p_outer} 
lists the same for {\bf G-2}. The results presented in Tables \ref{table:d_p_inner} and \ref{table:d_p_outer} illustrate that it is 
not possible to reproduce the observed phase relationship between the leading and the trailing peak with unaliased drifting $-$ it is 
too slow to move the spark pattern so much that the emission is correlated after $\sim$ 9$P_1$. Hence, the drifting is aliased for 
PSR B0818$-$41. For a particular geometry and a choice of $N_{sp}$, observed $P_5$ can be reproduced with certain choice of $D$ and $n$. 
Knowing $D$ and $n$, one can calculate the $P_3^t$ using Eqn. 1 of Paper 1. Since $P_4=N_{sp}\:P_3^t$; $P_4$ is automatically 
constrained for all the cases considered in Table \ref{table:d_p_inner} and  \ref{table:d_p_outer}. 
Cases for which $P_5$ is close to the observationally estimated value ($P_5= 9\pm 2$) are indicated by bold 
face in Table \ref{table:d_p_inner} and  \ref{table:d_p_outer}. 
For {\bf G-1}, $P_5= 9\pm 2$ is satisfied for the following choices,\\ 
(1) $N_{sp}=20$, 1st order alias ($n=2$, $k=1$), $D=19.05$\degr/$P_1$, $P_5=8.69P_1$, 
$P_4\sim18P_1\sim10$ s\\
(2) $N_{sp}=19$, 2nd order alias ($n=4$, $k=2$), $D=38.95$\degr/$P_1$, $P_5=7.40P_1$, 
$P_4\sim9P_1\sim5$ s\\
(3) $N_{sp}=18$, 2nd order alias ($n=4$, $k=2$), $D=41.05$\degr/$P_1$, $P_5=10.89P_1$, 
$P_4\sim9P_1\sim5$ s\\
For {\bf G-2}, $P_5= 9\pm 2$ is satisfied for the following choices,\\                                                                  
(1) $N_{sp}=20$, 1st order alias ($n=1$, $k=1$), $D=19.05$\degr/$P_1$, $P_5=8.74P_1$, 
$P_4\sim18P_1\sim10$ s\\
(2) $N_{sp}=17$, 2nd order alias ($n=3$, $k=2$), $D=43.40$\degr/$P_1$, $P_5=7.11P_1$, 
$P_4\sim9P_1\sim5$ s\\
According to the above analysis, 1st order aliasing with $D=19.05$\degr/$P_1$ and $N_{sp}=20$ give the best match of $P_5$ with observations, 
for both {\bf G-1} and {\bf G-2}, which is rather coincidental to us. Hence, it does not help to distinguish between {\bf G-1} and {\bf G-2}. 
We note that $P_4$ determined in this work is identical to the estimation in Paper 1 and supports the interesting result that the carousel 
rotation rate is the same as the measured $P_3^m$.

We simulate the radiation for each geometry for different combinations of $N_{sp}$ and $D$ (refer to Sect. \ref{sec:sim}). 
Next we correlate the emission under the leading and trailing peaks of the simulated single pulses at different pulse lags 
(as is done for the real data). $P_5$ determined from the simulated data is found to be identical to that calculated 
from Eqn. \ref{eqn4} (listed in Tables \ref{table:d_p_inner} and \ref{table:d_p_outer}). Hence, our methodology for 
determining $P_5$ is confirmed by simulations.

To summarise, the observed phase relationship between the peak emission from the leading and trailing outer region is an 
unique feature of PSR B0818$-$41 and acts as an additional information other than $P_1$, $P_2$, $P_3$ and $P_4$. Investigation 
of this unique phase relationship can solve the aliasing problem for this pulsar. It shows that drifting has to be aliased in 
PSR B0818$-$41. We provide a few combinations of feasible values of the $N_{sp}$, $D$ and $n$ that can produce the observed out 
of phase relation. The results from the corresponding simulations support our arguments. We propose, $P_4\:\sim\:18 P_1\:\sim$ 
10 s (for $n=2$ for {\bf G-1} and $n=1$ for {\bf G-2}). The attempt to determine $n$ and $P_4$ has been successful only for a 
few other pulsars : PSR B0809$+$74 ($P_4=165$ $P_1\sim200$ s, $n=2$, \cite{Leeuwen_etal}), PSR B0818$-$13 ($P_4=30$ s, 
\cite{Janssen_etal}), PSR B0826$-$34 ($P_4=14$ $P_1\sim25.9$ s, $n=2$, \cite{Gupta_etal}), PSR B0834$+$06 ($P_4=14.8$ $P_1\sim18.9$ s, 
\cite{Asgekar_etal}), PSR B0943+10 ($P_4=37$ $P_1\sim40$ s, $n=0$ \cite{Desh_etal}) and PSR B1857$-$26 ($P_4=147$ $P_1\sim89.9$ s, 
$n=0$, \cite{Mitra_etal_08}). Hence, PSR B0818$-$41 has the fastest known $P_4$, faster than PSR B0834$+$06.
\begin{table*}
\begin{minipage}{180mm}
\caption{Calculation of $P_5$ for {\bf G-1} (the inner geometry) for different spark arrangements}
\label{table:d_p_inner}
\begin{tabular}{l|c|c|c|c|c|c|c|c|c|c|c|c|c|c|c|c}
\hline
Drift rate     &Number of  &Azimuthal separation  &$\Delta\sigma/\Delta s$ & Apparent drift &Alias& $f_2$   &Pulse              \\
($\degr/P_1$)  &sparks     &between sparks        &                        &  ($D_{ap}$)    &order&         &offset            \\
           &($N_{sp}$) & ($\Delta s$)         &                  &                &($n, k$) &         & ($P_5$)  \\
           &           &                      &                  &                &         &         &                       \\\hline
0.95       & 19        & 18.95                & 3.09             &  0.95          & 0       & 0.10    & 17.86                           \\
           &           &                      &                  &                &         &         &                          \\\hline
20         & 19        & 18.95                & 3.09             &  1.05          & 2, 1    & 0.34    & 11.82                        \\
           &           &                      &                  &                &         &         &                        \\\hline
38.95      & 19        & 18.95                & 3.09             &  1.05          & 4, 2    & 0.50    & {\bf 7.40}                       \\
           &           &                      &                  &                &         &         &                       \\\hline
57.89      & 19        & 18.95                & 3.09             &  1.05          & 6, 3    & 0.83    & 3.15                          \\
           &           &                      &                  &                &         &         &                                 \\\hline
76.84      & 19        & 18.95                & 3.09             &  1.05          & 8, 4    & 1.08    & 17.46                          \\
           &           &                      &                  &                &         &         &                       \\\hline
21.05      & 18        & 20                   & 2.92             &  1.05          & 2, 1    & 1.18    & 15.53                        \\
           &           &                      &                  &                &         &         &                        \\\hline
41.05      & 18        & 20                   & 2.92             &  1.05          & 4, 2    & 1.43    & {\bf 10.89}                          \\
           &           &                      &                  &                &         &         &                          \\\hline
22.23      & 17        & 21.18                & 2.76             &  1.05          & 2, 1    & 1.02    & 19.72                        \\
           &           &                      &                  &                &         &         &                         \\\hline
43.40      & 17        & 21.18                & 2.76             &  1.05          & 4, 2    & 1.26    & 14.81                         \\
           &           &                      &                  &                &         &         &                            \\\hline
19.05      & 20        & 18                   & 3.25             &  1.05          & 2, 1    & 0.51    & {\bf 8.69}                         \\
           &           &                      &                  &                &         &         &                             \\\hline
37.05      & 20        & 18                   & 3.25             &  1.05          & 4, 2    & 0.75    & 4.22                        \\
           &           &                      &                  &                &         &         &                           \\\hline
18.17      & 21        & 17.14                & 3.41             &  1.05          & 2, 1    & 0.67    & 5.34                         \\
           &           &                      &                  &                &         &         &                          \\\hline
35.34      & 21        & 17.14                & 3.41             &  1.05          & 4, 2    & 0.92    & 1.36                        \\
           &           &                      &                  &                &         &         &                       \\\hline
\end{tabular}

$\dagger$: combinations producing, $P_5= 9\pm 2$ are indicated by bold face. 
\end{minipage}
\end{table*}
\begin{table*}
\begin{minipage}{180mm}
\caption{Calculation of $P_5$ for {\bf G-2} (the outer geometry) for different spark arrangements}
\label{table:d_p_outer}
\begin{tabular}{l|c|c|c|c|c|c|c|c|c|c|c|c|c|c|c|c}
\hline
Drift rate &Number of  &Azimuthal separation  &$\Delta\sigma/\Delta s$ & Apparent drift &Alias    & $f_2$   & Pulse        \\
($\degr/P_1$)  &sparks &between sparks        &                  &  ($D_{ap}$)    &order    &         & offset               \\
($D$)      &($N_{sp}$) & ($\Delta s$)         &                  &                &($n, k$) &         & ($P_5$)                \\
           &           &                      &                  &                &         &         &                       \\\hline
1.05       & 19        & 18.95                & 6.87             &  1.05          & 0       & 0.88    & 2.25                       \\
           &           &                      &                  &                &         &         &                      \\\hline
20         & 19        & 18.95                & 6.87             &  1.05          & 1, 1    & 1.13    & 15.71                         \\
           &           &                      &                  &                &         &         &                       \\\hline
38.95      & 19        & 18.95                & 6.87             &  1.05          & 3, 2    & 1.37    & 11.32                    \\
           &           &                      &                  &                &         &         &                \\\hline
57.89      & 19        & 18.95                & 6.87             &  1.05          & 5, 3    & 1.62    & 6.92                    \\
           &           &                      &                  &                &         &         &                \\\hline
76.84      & 19        & 18.95                & 6.87             &  1.05          & 7, 4    & 1.86    & 2.52                      \\
           &           &                      &                  &                &         &         &               \\\hline
21.05      & 18        & 20                   & 6.51             &  1.05          & 1, 1    & 0.76    & 4.48                    \\
           &           &                      &                  &                &         &         &                \\\hline
41.05      & 18        & 20                   & 6.51             &  1.05          & 3, 2    & 1.00    & 18.83                    \\
           &           &                      &                  &                &         &         &                \\\hline
22.23      & 17        & 21.18                & 6.14             &  1.05          & 1, 1    & 0.40    & 12.02                          \\
           &           &                      &                  &                &         &         &               \\\hline
43.40      & 17        & 21.18                & 6.14             &  1.05          & 3, 2    & 0.65    & {\bf 7.11}                    \\
           &           &                      &                  &                &         &         &               \\\hline
19.05      & 20        & 18                   & 7.23             &  1.05          & 1, 1    & 0.49    & {\bf 8.74}                  \\
           &           &                      &                  &                &         &         &         \\\hline
37.05      & 20        & 18                   & 7.23             &  1.05          & 3, 2    & 0.73    & 4.56                   \\
           &           &                      &                  &                &         &         &         \\\hline
18.17      & 21        & 17.14                & 7.59             &  1.05          & 1, 1    & 0.85    & 2.43                  \\
           &           &                      &                  &                &         &         &         \\\hline
35.34      & 21        & 17.14                & 7.59             &  1.05          & 3, 2    & 1.09    & 14.73                   \\
           &           &                      &                  &                &         &         &         \\\hline
\end{tabular}

$\dagger$: combinations producing, $P_5= 9\pm 2$ are indicated by bold face.
\end{minipage}
\end{table*}
\subsection{Phase locked relation (PLR) between inner and outer ring}
From the gray scale plot of the single pulses (e.g. Fig. \ref{sp_325}) it is evident that inner and outer drift regions 
are locked in phase : inner drift bands start from almost halfway between two drift bands in the leading outer region 
and always remain connected with the drift bands from the trailing outer region. This constant phase relation does not 
get perturbed by nulling or irregular drifting and is maintained even during the sequences of irregular drifting as well 
as after nulling. The PLR implies that the emission from the inner and the outer rings are always in fixed phase. Furthermore, 
nulling and changing drift rates are observed to be simultaneous in both the rings. Interestingly, a 180\degr phase shift 
between the spark locations on the inner and the outer rings is needed in the simulation to match the PLR of the observed 
drift pattern of PSR B0818$-$41. Such an arrangement of sparks corresponds to a maximally packed structure on the polar cap. 
Spark discharges occur in every place where the potential drop is high enough to ignite and develop pair production avalanche 
\citep{Ruderman_etal}, so sparks populate the polar cap as densely as possible. At the same time, such an arrangement of 
sparks should be quasi rigid, a property which is also suggested by the data.
 
PSR B0826$-$34 is another wide profile pulsar for which presence of simultaneous multiple drift regions are observed 
(\cite{Biggs_etal}, \cite{Esamdin_etal} and \cite{Bhattacharyya_etal_08}). For this pulsar the main pulse and the inter 
pulse emission are interpreted to be coming from two concentric rings of emission (\cite{Gupta_etal} and \cite{Esamdin_etal}). 
Fig. 9 of \cite{Bhattacharyya_etal_08} presents the gray scale plot of the single pulses from PSR B0826$-$34 at 610 MHz. 
This plot illustrates simultaneous drifting in the main pulse ($\sim$ 5 to 6 drift bands) and inter pulse ($\sim$ 3 to 4 drift 
bands) region. The drift bands in these regions are locked in phase implying that the emission from the inner and the outer 
rings are in phase. \cite{Esamdin_etal} applied the method of phase tracking to the single pulse data and detected in total 
13 drift bands in the pulse window. PLR between the inner and outer rings is evident from their results (Fig. 6 of 
\cite{Esamdin_etal}). For PSR B0826$-$34 we observe frequent nulling and changes of drift rates which are simultaneous for 
both the inner and outer rings. In addition, \cite{Bhattacharyya_etal_08} reported significant correlation between total 
energy of the main pulse and the inter pulse. They found that with increasing pulse lag this correlation follows similar 
trend as the autocorrelation of total energy of the main pulse or the inter pulse (Fig. 7 of \cite{Bhattacharyya_etal_08}). 
The only other pulsar for which simultaneous occurrence of multiple drift regions and phase relation between the drift regions 
are observed is PSR B0815+09. But the mirrored drift bands observed for PSR B0815$+$09 can not be explained within the 
\cite{Ruderman_etal} model.

Some pulsars $-$ e.g. PSR B1237$+$25, B1857$-$26 $-$ clearly shows multiple components in their profile (mostly 
five components, denoted by "M"). \cite{Rankin_93} have recognised these pulsars as a distinct observational class 
of objects, where the emission is coming from the core and two concentric conal rings. No signature of drifting is 
detected in the inner ring for these pulsars, possibly because of the fact that the inner ring is too weak to detect 
any single pulse behaviour. For example, in PSR B1237$+$25 drifting is observed in the outer ring with fluctuation 
feature at $P_3=2.8P_1$ \citep{Zuzana_etal}. In Fig. 6 of \cite{Zuzana_etal} we observe a faint fluctuation feature 
around $P_3=2.8P_1$ corresponding to the inner ring.  

To summarise, signatures of drifting from more than one ring are observed for only a few pulsars and for all of 
those pulsars, emission from inner and outer rings are locked in phase. No counter example is observed. This requires 
common drift rate in the outer and inner rings, implying that emission in the two rings are not independent, and that 
the conditions responsible for drifting are similar in both rings. This demands that the conditions are similar everywhere 
in the magnetosphere, which can in turn put constraints on the theoretical models attempting to explain the pulsar 
emission mechanism. Our finding of PLR between the emission from the inner and the outer rings can be explained 
considering the pulsar emission as a pan magnetospheric phenomenon. 

According to \cite{Gil_etal_00}, at any given time, the polar cap is populated as densely as possible. They consider, 
characteristic dimension of a spark and typical distance between sparks being equal to polar cap height h. They do not 
distinguish between the radial and the azimuthal directions, which may favor similar conditions (electric field and 
magnetic field) for both the rings of emission, leading to pan magnetospheric conditions across the polar gap. 
\cite{Wright_03} proposed an alternative model for the pulsar magnetosphere which considers the magnetosphere as an 
integral whole, and suggests that the inner and outer cones found in integrated profiles reflect the two intersections 
of the null surface with the light cylinder and the co-rotating dead zones respectively. They consider the pulsar radio 
emission phenomenon as a global pan magnetospheric phenomenon. Our results are consistent with these kinds of models.

\section{Summary}                      \label{sec:Summary}
Following are the interesting new results from our investigation of subpulse drifting and polarization properties of PSR 
B0818$-$41 at multiple frequencies:\\
(1) We estimate the mean flux of PSR B0818$-$41 at 5 different frequencies and show that the spectrum flattens at frequencies 
lower than 325 MHz (at 244 or 157 MHz), providing indication of a low frequency turn-over. \\
(2) We present the average profiles at five different frequencies. The evolution of profile width with frequency follows radius 
to frequency mapping. \\
(3) Significant linear polarization is observed at 325, 610 and 1060 MHz. Linear polarization profiles follow similar trend as 
the total intensity profile and get depolarized at both edges of the profile. This can be explained by the observed orthogonal 
polarization mode jump at the edges of the profile. Polarization angle sweep across the pulse profile evolves remarkably with 
frequency (at 325, 610 and 1060 MHz), which is not generally observed in other pulsars. Though very less circular polarization 
without any signature of changing handedness is observed at 325 and 610 MHz, circular polarization changes sign at the middle 
of the pulse profile at 1060 MHz. \\
(4) In addition to the remarkable subpulse drift observed at 325 MHz, we report subpulse drifting at 244, and 610 MHz. At 
244 MHz subpulse drifting is observed only in the leading and trailing outer regions and not in the inner region. Though the 
drift bands are weaker, subpulse drifting is observed in both inner and outer region at 610 MHz. Frequent changes of drift 
rates, including transitions of drift rates from negative to stationary and vice verse, are observed for this pulsar. Such 
transitions of drift rates appear to have some connection with nulling. $P_2^m$, $P_3^m$ and $\Delta\Phi_s$ are determined 
for the inner and the outer drift regions. Though $P_3^m$ is observed to be the same for the inner and outer drift regions, 
$P_2^m$ and $\Delta\Phi_s$ are different for different drift regions. \\
(5) We report that the peaks of the emission from the trailing and leading outer regions, as a function of the pulse number, are 
offset by a constant interval, $P_5$$\sim$9$P_1$. We also report a phase locked relation (PLR) between the inner and outer 
drift regions for PSR B0818$-$41.\\

Interpretations of the analysis results, aided by the simulations, lead to following new outcomes,\\
(1) Based on the frequency evolution of the average profile, observed PA swing and results from subpulse drifting, we converged 
on two possible choices of emission geometry: {\bf G-1} ($\alpha=11$\degr and $\beta=-5.4$\degr) and  {\bf G-2} ($\alpha=175.4$\degr 
and $\beta=-6.9$\degr). Pulsar radiation pattern simulated with both the geometries reproduces the observed features, except for 
some differences. Rather surprisingly, all other observed properties are found to be satisfied for both these geometries. \\
(2) We explain the observed subpulse drifting and transitions of apparent drift rates from negative to stationary as being due 
to aliasing. The frequent transitions from negative to stationary drift rates observed for PSR B0818$-$41 can be understood by 
slight changes in polar cap temperature leading to change in the drift rate so that $P_3^t=P_1$ is satisfied and longitude 
stationary drifting is observed. \\
(3) The absence of the inner drift region at 244 MHz is interpreted as, the LOS missing the emission from the inner ring. This is 
supported by the results from the simulations.\\
(4) A new technique is introduced in this paper for resolving aliasing, using the constant offset ($P_5\sim9P_1$) between the peak 
emission from the leading and trailing outer regions. From the result of this technique, we propose that the subpulse drifting for 
PSR B0818$-$41 is most likely first order aliased, and the corresponding carousel rotation period $P_4=10$ s. This implies that 
PSR B0818$-$41 has the fastest known carousel.\\ 
(5) 180\degr phase shift between the spark locations on the inner and the outer rings is needed in the simulation to match the 
observed PLR of the drift pattern, which implies that the sparks that are maximally packed on the polar cap.\\
(6) For all pulsars for which we know drifting from more than one ring (e.g. PSR B0826$-$34 and possibly PSR B0815$+$09), the drift 
pattern in the inner and outer rings are always phase locked. This could be a significant constraint for the theoretical models of 
pulsar radio emission, and favors an emission model that considers pulsar radio emission as a pan magnetospheric phenomenon.

Hence, the results from the observations, simulations and the follow up interpretations, enable fair amount of insight towards 
the emission properties of the unique drifting pulsar B0818$-$41. This study will also have significant contribution to 
constrain the theoretical models explaining the possible emission mechanism of pulsars.\\

\noindent{\large\bf Acknowledgments :}
We thank the staff of the GMRT for help with the observations. The GMRT is run by the National Centre for Radio Astrophysics of the 
Tata Institute of Fundamental Research. We thank D. Mitra for polarization data analysis programs and insightful discussions 
about the emission geometry, R. Thomas for discussion about polarization, G. Wright for helpful discussions regarding theoretical 
models of pulsar radio radiation, M. Sendyk for help in simulations and R. Smits for the fluctuation spectrum analysis routine. J. G. 
acknowledge a partial support of Poslish Grant N N203 2738 33.

\label{lastpage}
\end{document}